\documentclass[acmtog,timestamp]{acmart} 

\usepackage{amsbsy}
\usepackage{ amssymb }
\usepackage{subfig}
\usepackage{graphicx}
\usepackage{enumitem}
\setlist{nolistsep,leftmargin=*}
\usepackage{bm}
\usepackage{amsfonts}
\usepackage{mathtools}
\usepackage{dsfont}
\usepackage{algorithmicx}
\usepackage{algorithm}
\usepackage{algpseudocode}
\usepackage[font=small,skip=3pt]{caption}
\usepackage{multirow}
\usepackage{makecell}
\usepackage{hyperref}
\usepackage{soul}
\algnotext{EndFor}
\algnotext{EndIf}
\algnotext{EndWhile}
\hypersetup{
    colorlinks=true,
    linkcolor=blue,      
    urlcolor=blue,
}

\algrenewcommand\algorithmicindent{0.8em}%

\newcommand{\nix}[1]{}
\newcommand{\CS}{\emph{color sail}}
\newcommand{\CSs}{\emph{color sails}}

\newcommand{\edits}[1]{\textcolor{blue}{#1}}

\renewcommand{\edits}[1]{#1}
\renewcommand{\st}[1]{}

\newcommand{\patentnote}[1]{\textcolor{magenta}{ {\bf NOTE: } #1}}
\renewcommand{\patentnote}[1]{}

\setcopyright{none}

\acmConference[SIGGRAPH 2018]{SIGGRAPH 2018}{August 2018}{Vancouver}
\acmYear{2018}

\begin{document}

\title{Color Sails: Discrete-Continuous Palettes for Deep Color Exploration}

\author{Maria Shugrina}
 \affiliation{%
   \institution{University of Toronto}
 }

\author{Amlan Kar}
 \affiliation{%
   \institution{University of Toronto}
 }

\author{Karan Singh}
 \affiliation{%
   \institution{University of Toronto}
 }
 
\author{Sanja Fidler}
 \affiliation{%
   \institution{University of Toronto}
 }

\citestyle{acmauthoryear}
\setcitestyle{square}

\begin{abstract}
We present \emph{color sails}, a discrete-continuous color gamut representation that extends the color gradient analogy to three dimensions and allows interactive control of the color blending behavior. Our representation models a wide variety of color distributions in a compact manner, and lends itself to applications such as color exploration for graphic design, illustration and similar fields. \edits{We propose a Neural Network that can fit a color sail to any image. Then, the user can adjust color sail parameters to change the base colors, their blending behavior and the number of colors, exploring a wide range of options for the original design (Fig.~\ref{fig:teaser}). In addition, we propose a Deep Learning model 
that learns to automatically segment an image into \st{semantically meaningful} color-compatible alpha masks, each equipped with its own color sail. This allows targeted color exploration by either editing their corresponding color sails or using standard software packages. Our model is trained on a custom diverse dataset of art and design. We provide both quantitative evaluations, and a user study, demonstrating \st{both} the effectiveness of color sail interaction. \st{, and our automatic methods.}
Interactive demos are available at \textbf{\href{http://www.colorsails.com}{www.colorsails.com}}}.



\end{abstract}

%
%


\begin{teaserfigure}
  \includegraphics[width=\textwidth]{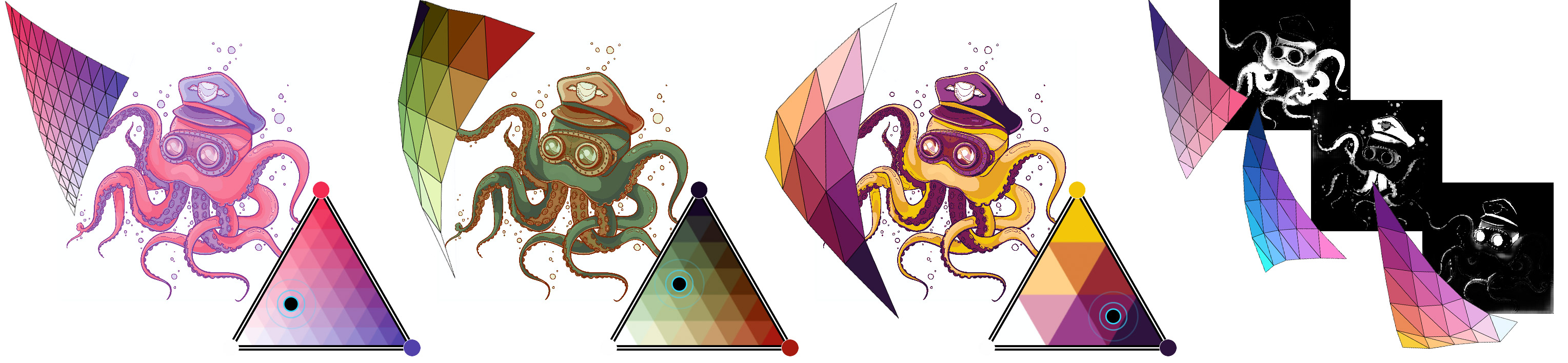}
  \caption{Color sail palettes allow interactive control over any design's colors with just a few knobs. An artist can control not just the base palette colors, but its non-linear blending behavior and discretization level. We propose an automatic method to rig any existing image with a collection of region-based color sails, allowing fast interactive exploration of color relationships in an image.}
  \label{fig:teaser}
\end{teaserfigure}

\maketitle

\section{Introduction}
Color is one of the key components in art and design. A director's choice of a color theme can set the mood for an animated film, and the accent color selected for a website can radically change the way it is perceived. While it is easy to agree that a "color theme" is vital to any visual creation, there exists no unified color theme representation that can span various domains, such as painting, illustration, graphic and user interface design, where both discrete and continuous color distributions may be required. We propose \emph{color sails}, a discrete-continuous color gamut representation that is compact, interactive and naturally integrates into deep learning architectures for color exploration.



The utility of a unified color representation capable of modeling a variety of color distributions is easy to see through the example of discrete color palettes. There has been a large body of research that relies on discrete color palettes, typically comprised of 5-10 colors, e.g.\ for recoloring, color theme extraction, and machine learning of color aesthetics. Discrete color palettes have also become popular outside of the academic community and are actively created, shared and rated using commercial tools. This suggests a need to represent color in a way that is decoupled from any artwork or design. While discrete color palettes may be suitable for some design domains, such as user interface design, they are a poor representation for much larger, continuous color distributions found in most other art and design work, such as illustration, graphic design and photography. Despite these limitations, discrete color palette's simplicity and the ease of its construction have made it a popular tool for artists and researchers alike. Our representation combines this kind of simplicity with representational power and versatility.



Each \emph{color sail} models a patch of RGB (or any other color space) with 4 interactive controls: three colors defining the color interpolation space, and a fourth control that modifies the shape of the gradient (\S\ref{sec:rep}). This allows "bending" the color distribution by putting "wind" into the sail, approximating a variety of non-linear color blending behaviors. In addition, a color sail has a parameter for the level of discretization, making it possible to model a range of color palettes from very coarse to visually continuous. 
We aim to automatically infer a small set of color sails for an image, and map each to a meaningful image region. This allows users to easily recolor images by interacting with the sail controls. Unlike past work~\cite{shugrina2017playful} which is typically limited to artwork created with custom palette interfaces, we aim to work  with any design.

The contributions of our work are as follows. First, we propose \CSs{}, a compact and interactive color gamut representation, able to model both discrete and continuous distributions (\S\ref{sec:rep}). We motivate the application of \CSs{} for interactive color exploration in existing designs (\S\ref{sec:rig}). Second, we formulate \CS{} as a 
	fully differentiable module compatible with deep learning architecture, and propose a neural network that fits a color sail to a color distribution by learning from blending behavior in real artwork image patches (\S\ref{sec:pg}). Third, to model more varied color distributions, 
	we propose a deep learning method that learns to segment images into soft regions, each fitted with its own \CS{} (\S\ref{sec:learning}). This approach is a step toward semantics-aware color editing applications, and we hope that our color representation will be used for other work in this area. Our target domain is human-generated designs, visualization and digital art, necessitating the use of a custom training data set (\S\ref{sec:data}). We demonstrate the results of our automatic methods and their editing capabilities in \S\ref{sec:results}, and conclude with a user study of our interactive system (\S\ref{sec:user_study}) and discussion (\S\ref{sec:discuss}).


\section{Related Work}\label{sec:prev}
\textbf{Discrete color palettes}: Discrete color palettes, consisting of a small (typically 5) number of solid color swatches, have become a ubiquitous representation of color combinations in design applications. Their simple representation makes these discrete palettes, or themes, easy to construct and share, e.g. using a platform such as Adobe Color CC \cite{kuler}, or \cite{colourlovers}. Perhaps due to this simplicity, discrete color palettes have also received substantial attention from the research community. A number of approaches have been proposed for extracting discrete color palettes from images \cite{odonovan2011color,lin2013modeling,chang2015palette}, and also for using these palettes for image recoloring \cite{wang2010data,chen2014sparse,chang2015palette} or colorization \cite{lin2013probabilistic}.  In addition, crowd sourcing user preferences for this canonical color theme representation inspired work on computational aesthetics of palettes \cite{odonovan2011color,odonovan2014collab}, and machine learning approaches to palette extraction and generation \cite{colormind}. Other methods for discrete color palette exploration have been proposed, e.g. a constraint-based exploration method of \cite{mellado2017constrained}. However, color distributions of most man-made images are not well-modeled by small discrete sets of colors, limiting the applicability of discrete color palettes. 

\textbf{Continuous color palettes}: Unlike discrete color palettes, color gamuts consisting of a large number of colors have no canonical representation. Perhaps for this reason they are harder to construct, analyze and edit, and have received less attention from the research community than the discrete color palettes. \cite{nguyen2015data} fit color manifolds to collections of images, but their color manifold representation is not editable once constructed and requires a large amount of data to approximate. On the other hand \cite{shugrina2017playful} propose a compact continuous palette representation, focusing on its interactive qualities, but their approach is not well-suited to the discrete setting, cannot model non-linear blending behavior and is not structured for data-driven analysis. Our goal is to provide a structured representation that can span a range from discrete to continuous color palettes, while modeling a variety of color blending behaviors and allowing interactive editing.

\textbf{Color blending:} 
Digital color interpolation is a direct consequence of the way color is represented. Linear interpolation in CIELAB~\cite{de2001improvement}, HSV~\cite{smith1978color}, HWB~\cite{smith1996hwb} and RGB results in vastly different gradients. Typically, graphic design software exposes only RGB interpolation due to its simplicity. However, RGB blending can result in grayish tones when transitioning from one vibrant color to another (e.g. blue to yellow). The team behind Paper53 App \shortcite{paper53} hand-tuned the blending behavior of paint pairs to look more pleasing~\cite{paper53tech}. Another possibility is to model the physical blending of paints with the Kubelka-Munk equation~\cite{Hecht1983}, e.g.\ to simulate watercolor \cite{curtis1997computer}. It is also possible to take a data-driven approach and learn blending and compositing behavior of a particular set of paints from examples~\cite{lu2014realpigment}.
Rather than settle on a particular alternative to the standard linear RGB interpolation we propose a way to model a variety of blending behaviors with the color sail's wind parameter. Our system exposes the control of this parameter to the artist, but it would also be possible to restrict it to better match a particular medium in a more targeted application.

\textbf{Recoloring and Colorization:}
Chang et al.~\shortcite{chang2015palette} extract a \emph{discrete} color palette from an image and allow changing it to recolor the image. Because their palette is discrete it allows users only indirect control over the resulting colors; the transformations are also applied to the image globally. In contrast, we use color sails to model the entire color distribution, thus allowing the user to shift emphasis from one color to the other, or to adjust color blending behavior. In addition, we allow finer control by fitting multiple palettes to regions of the image. Shugrina at el.~\shortcite{shugrina2017playful} allow the user to recolor the painting by changing a \emph{continuous} palette, but their method applies only to artwork created with their palette interface. In contrast, our method is designed to work with any design. Related to recoloring are user-guided systems for grayscale image colorization~\cite{paintchainer,zhang2017coloriz,sangkloy2016scribbler}. While these systems effectively learn image semantics for colorization, they are not designed to edit an image that already contains color. With our interactive recoloring system, we address a common use-case of exploring color choices in a design where color relationships have already been defined by the creator. 

\textbf{Photo Segmentation and Editing:} 
Computer vision field has addressed the problem of segmenting a photograph into regions in a variety of contexts~\cite{chen2016deeplab,qi20173d}, including portrait segmentation for image editing~\cite{shen2016automatic}. While these methods leverage data to infer image semantics, it is not their goal to allow direct color-based manipulation of the segmented regions. There is a significant body of work on color segmentation with the aim to split an image into soft or hard regions, e.g.\ most recently~\cite{tan2017decomposing,aksoy2017unmixing}. While these works provide useful layer segmentation of the image for editing, they do not explicitly encode color relationships within image regions because they work with constant color mats. Instead, we model color distributions in a more flexible way. This, together with a deep learning architecture capable of learning from image data, allows us to segment images in a way that does not rely solely on color and position cues, a property that can be useful for more semantic aware image editing. \edits {In a concurrent work to ours, \cite{aksoy2018semantic} use both color cues and deep semantic features for image segmentation. While their matting approach is related, it does not provide any new color editing capabilities, as does our work. Additionally, the deep learning architecture in \cite{aksoy2018semantic} requires a semantic segmentation dataset, limiting it to the domain of photography. Instead, our approach is unsupervised, applicable to a wide range of visual domains (See \S\ref{sec:data}) and takes special care to model color relationships in each region to facilitate editing.}


\textbf{Machine Learning for Image Manipulation:} 
Recently, a number of deep learning methods for image manipulation have been proposed. Methods such as~\cite{neuralstyle} allow users to change the style of the photograph, change its season, or swap out horses with zebras~\cite{cycleGAN}. In~\cite{iGAN}, the authors allow users to generate images on the fly in an interactive manner. In our work,  we aim to use machine learning in order to produce a set of user controls (alpha masks and color sail palettes) that give users \emph{direct} control to easily recolor their images.

\section{Overview}\label{sec:over}
\edits{We dub our color gamut representation \emph{color sail} due to its shape in the 3D color space. To achieve a variety of smooth non-linear blending behaviors, we model interpolation with a Bezier triangle that users can deform by applying \emph{wind}. We argue that this representation can be a useful tool in its own right, but our differentiable mathematical formulation also makes it possible to use \emph{color sails} with Stochastic Gradient Descent in a Deep Learning architecture. This property can make color sails useful for a number of applications in color, of which we define one.}

\edits{Governed by only 6 parameters, color sails are well-suited for interactive editing of color distributions not just in isolation, but also within existing designs. If the color of each pixel or higher level primitive such as a curve in a design is indexed by a given color sail, then the artist can manipulate the color sail to recolor the artwork. We call this image to color sail mapping a \emph{color sail rig}. Unlike prior work \cite{shugrina2017playful}, we propose an unsupervised Deep Learning method to automatically rig \emph{any existing design} with a color sail rig. This allows designers rapid interactive exploration of color alternatives (Fig.~\ref{fig:one_sail_edit}, Fig.~\ref{fig:alpha_res}), something that is practically useful according to our user study.}

\edits{First, we detail the color sail formulation (\S\ref{sec:rep}) and the color sail rig (\S\ref{sec:rig}). We then introduce our Neural Network that learns to generate these rigs automatically (\S\ref{sec:riglearn}), allowing users rapid color exploration of any design (\S\ref{sec:ux}). Results (\S\ref{sec:results}) and user study (\S\ref{sec:user_study}) follow.}

\section{Color Sail Representation}\label{sec:rep}
We propose \emph{color sail}, a versatile, compact, interactive representation for a variety of color distributions, from discrete to continuous. 

\subsection{Requirements}

\begin{figure}[h!]
\centering
\subfloat[linear RGB vs. acrylic paints]{%
  \includegraphics[width=0.49\columnwidth]{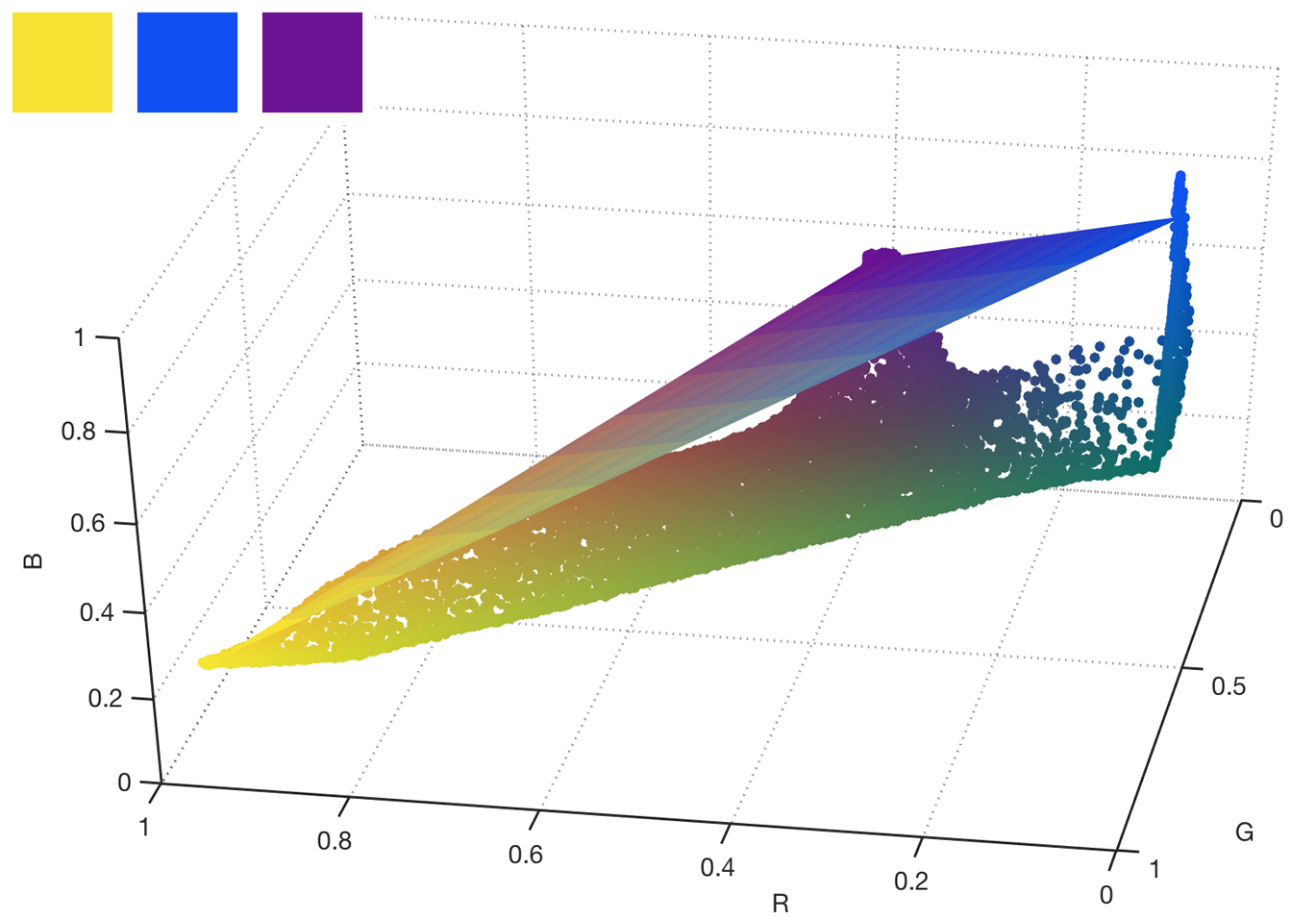}%
  }
\subfloat[color sail blending, varying $w$]{%
  \includegraphics[width=0.49\columnwidth]{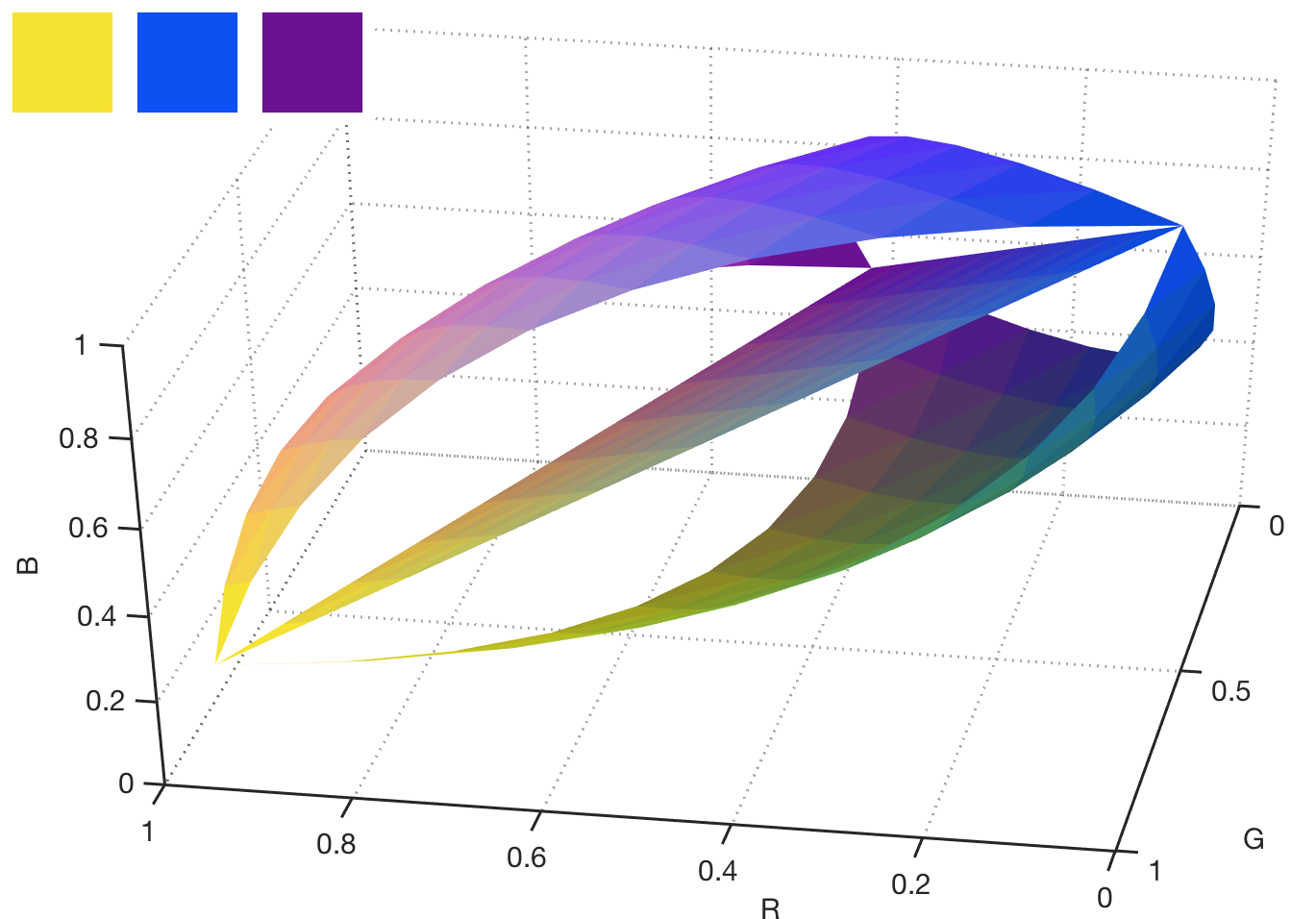}%
  }\\
  \subfloat[effect of wind on color]{%
  \includegraphics[width=0.99\columnwidth]{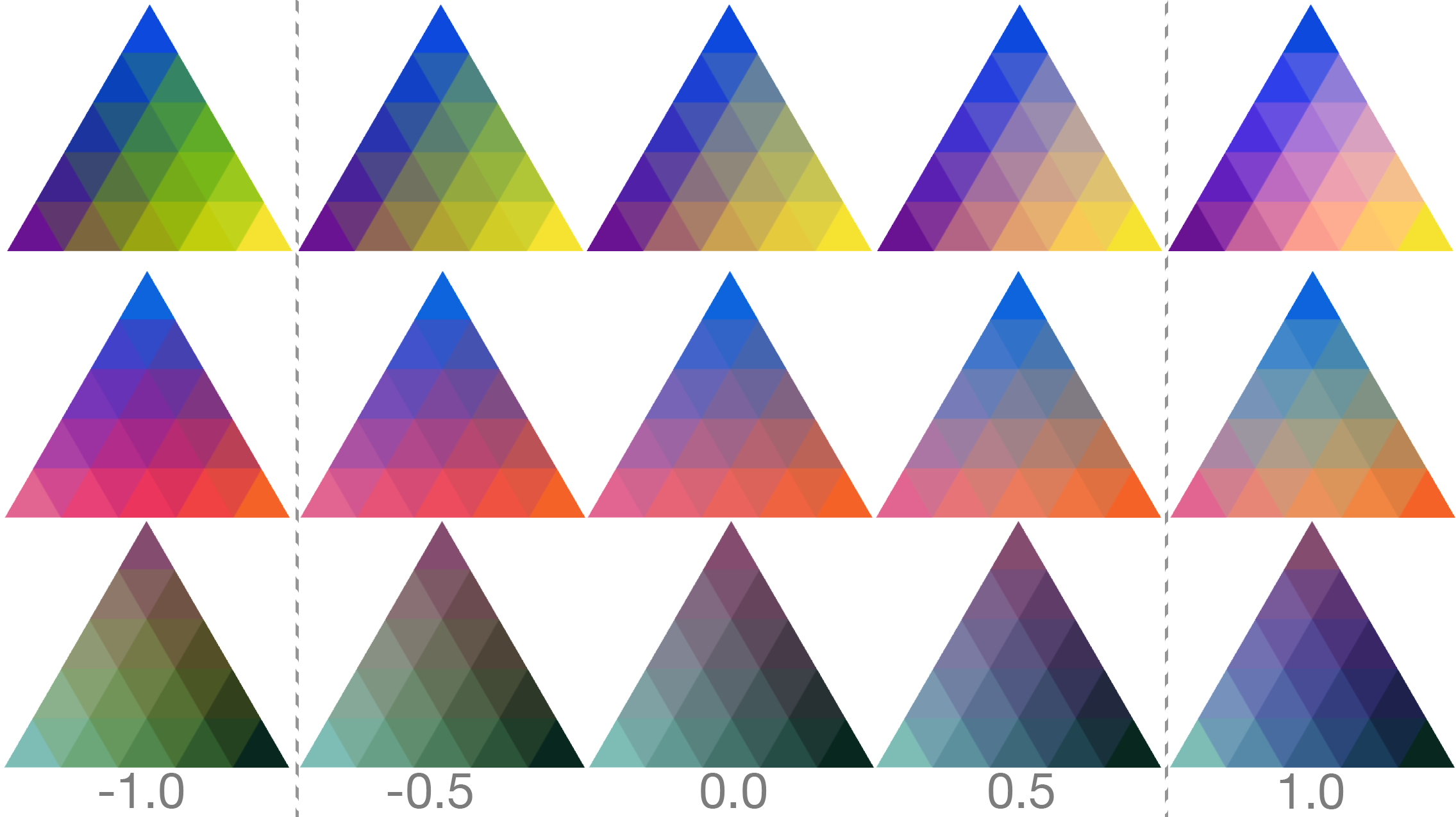}%
  }
\caption{\textbf{The Shape of Blending:} Linear RGB interpolation (plane in a) is a poor fit for many color blending scenarios, e.g. blending of real acrylic paints (manifold in a) modeled with Kubelka-Munk equation~\cite{Hecht1983} and paint coefficients from ~\cite{okumura2005developing}. Color sails offer a versatile way to model a variety of smooth non-linear blending behaviors (b). }
\label{fig:km}
\end{figure}

Our direct goal is to extend the appealing properties of discrete palettes with a representation that can model much more varied color distributions. Therefore, by direct analogy with discrete palettes, this representation must be:
\begin{enumerate}
\item \textbf{Sparse}: be comprised of few degrees of freedom
\item \textbf{Interactive}: easy to construct and edit interactively
\item \textbf{Learnable}: well-structured for learning and statistical analysis
\item \textbf{Versatile}: can model a wide array of color distributions	
\end{enumerate}
Here, requirements 1-3 are shared with discrete color palettes, and requirement 4 extends it to other domains. In particular, requirement 4 implies ability to model both discrete color gamuts, comprised of a handful of colors, and more complex color manifolds, common in art and design. While requirement 3 is hard to pin down exactly, we show that our representation is well-suited for machine learning by demonstrating two direct applications in \S\ref{sec:pg} and \S\ref{sec:learning}.

\subsection{Anatomy of a Color Sail}\label{ssec:anatomy}

\begin{figure}[ht]
\centering
\subfloat[naive]{%
  \includegraphics[width=0.33\columnwidth]{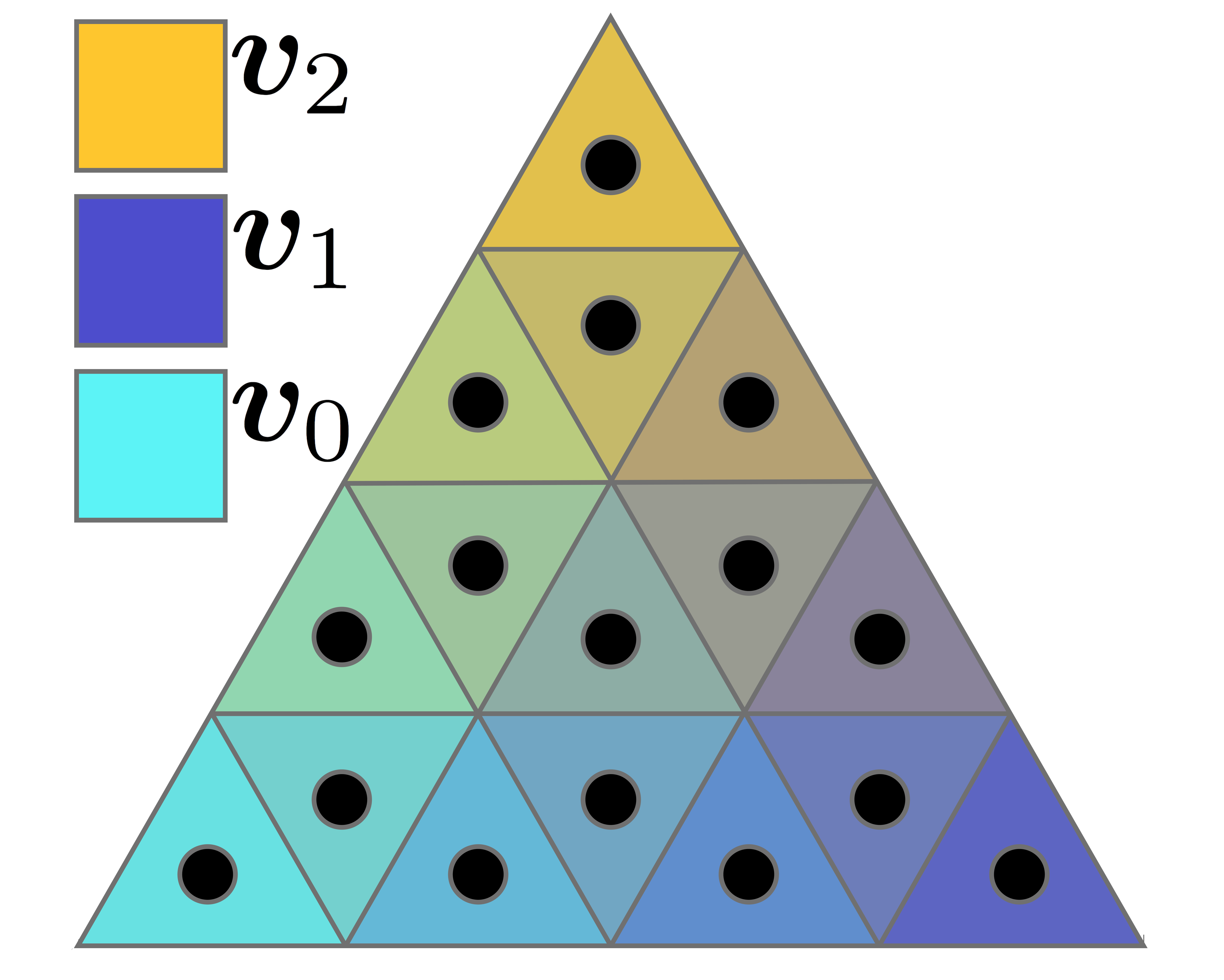}%
  }
\subfloat[ours]{%
  \includegraphics[width=0.33\columnwidth]{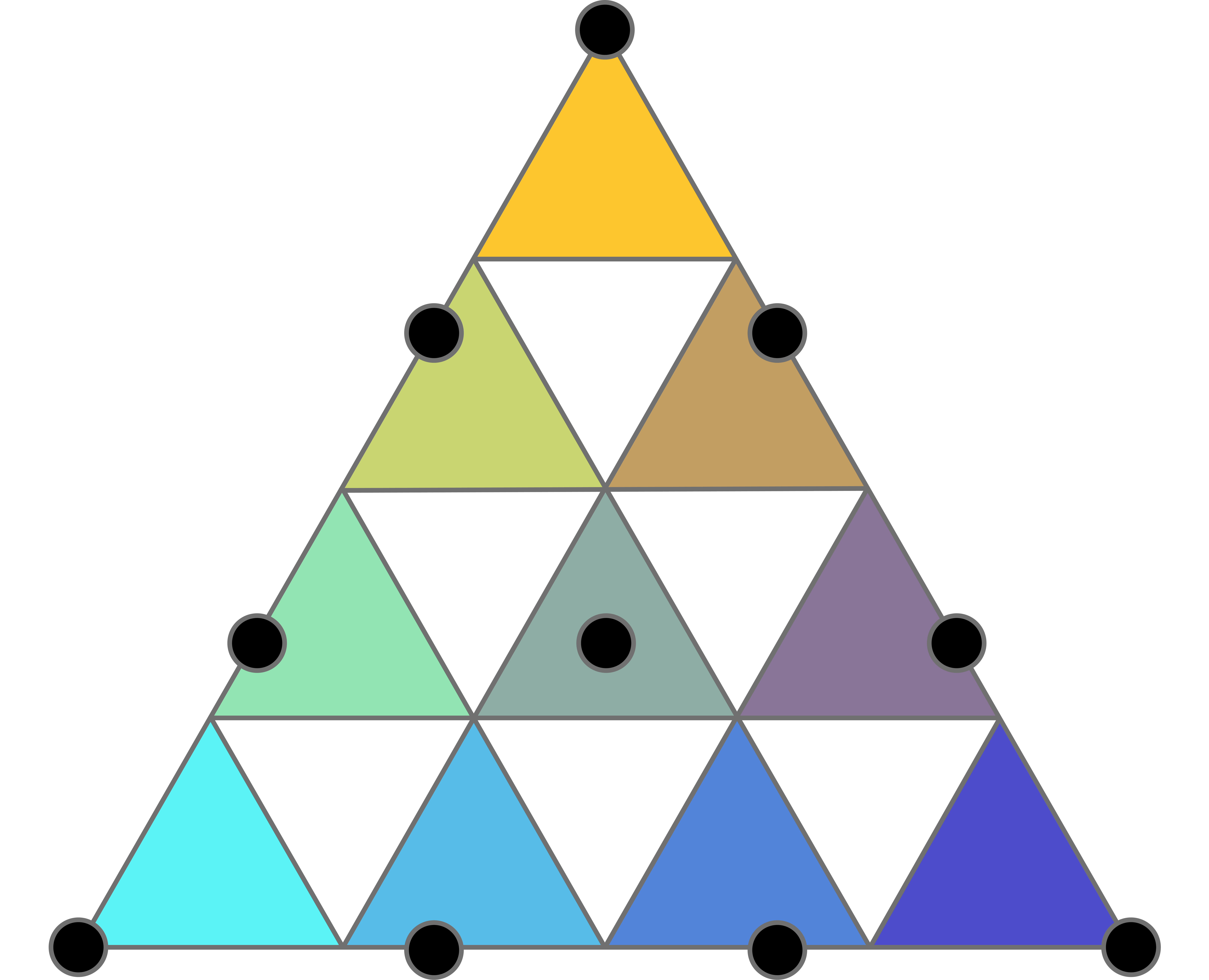}%
  }      
\subfloat[ours, expanded]{%
  \includegraphics[width=0.33\columnwidth]{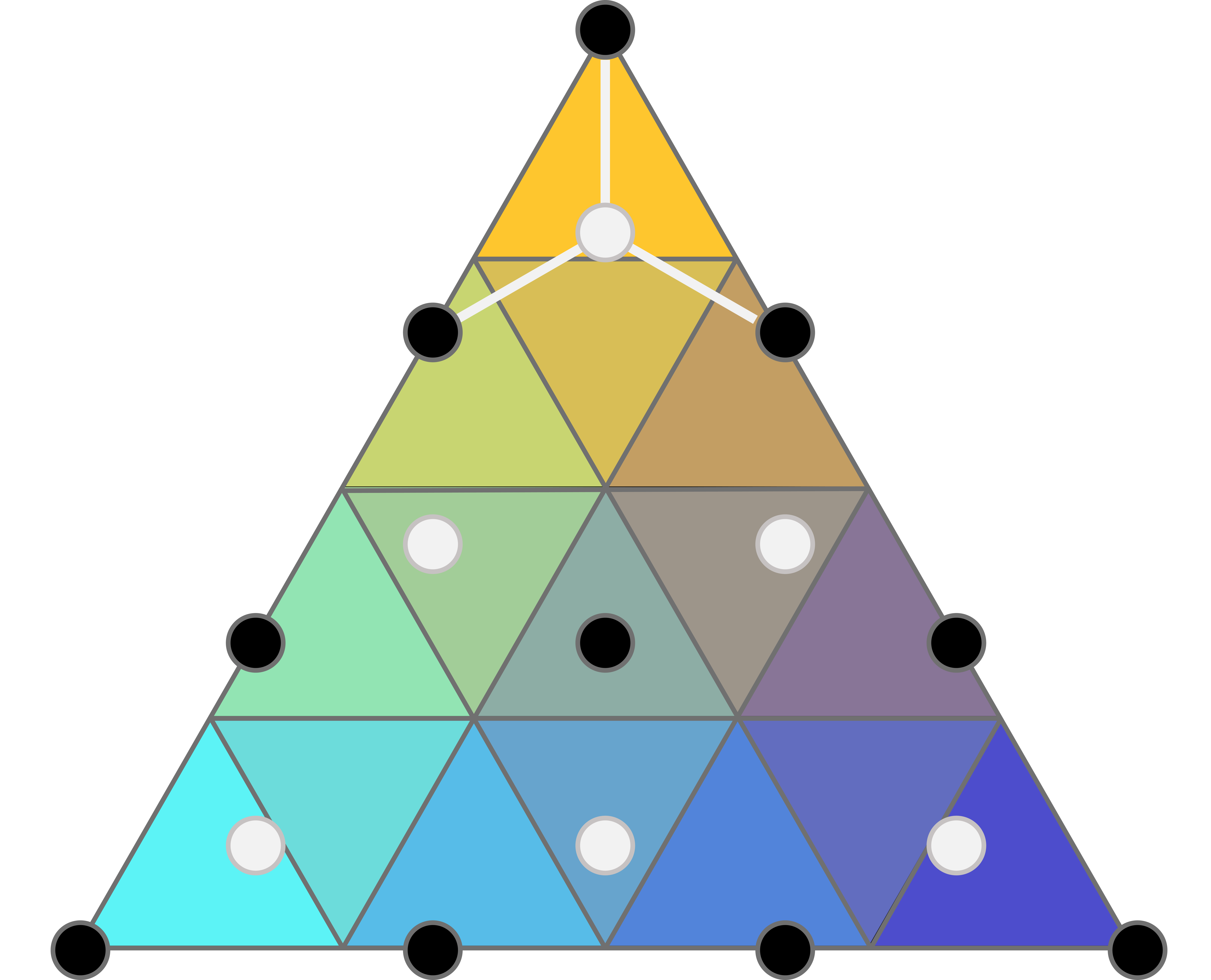}%
  } \\
  \subfloat[patchwork $s=2$]{%
  \includegraphics[width=0.33\columnwidth]{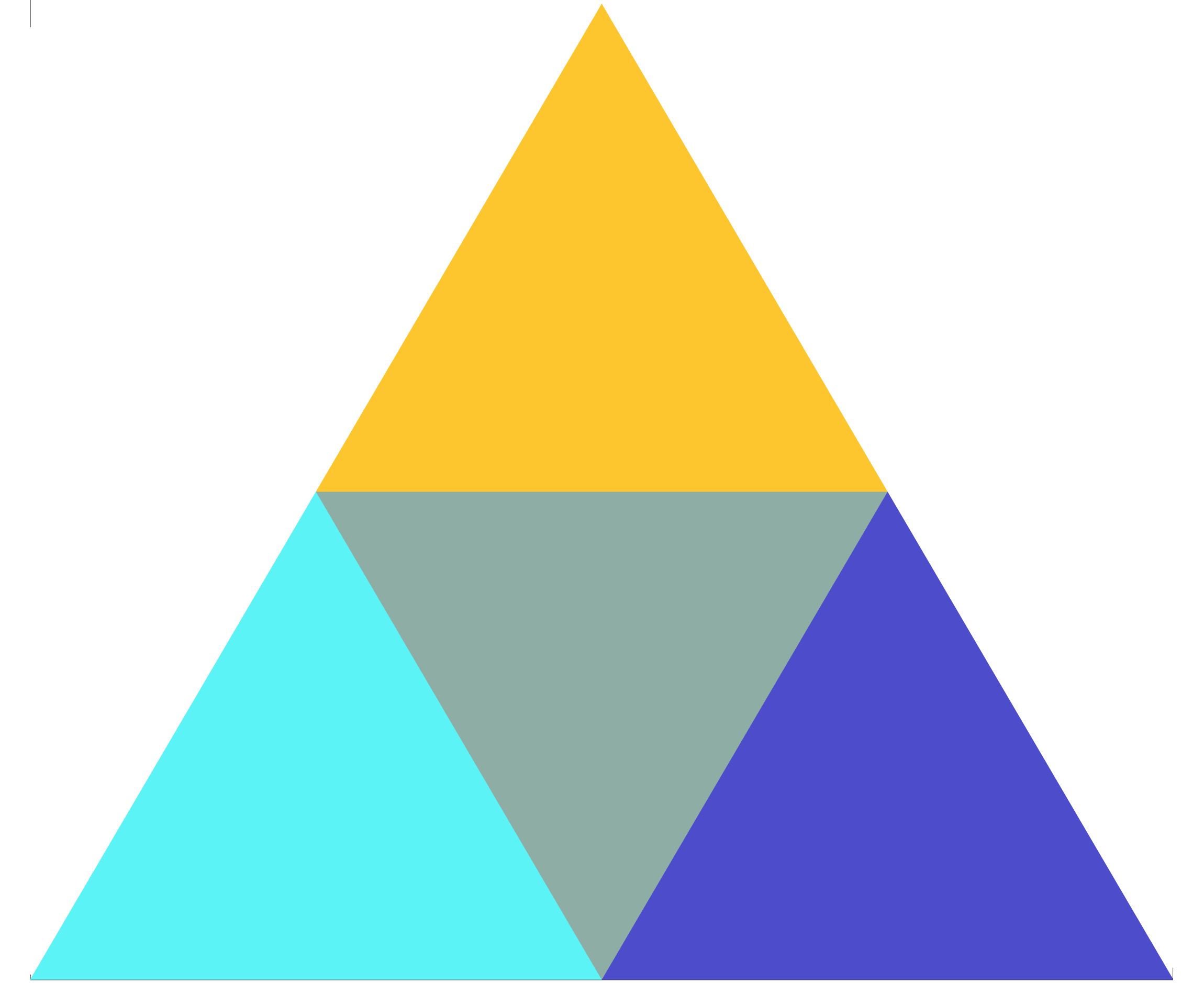}%
  }
\subfloat[patchwork $s=5$]{%
  \includegraphics[width=0.33\columnwidth]{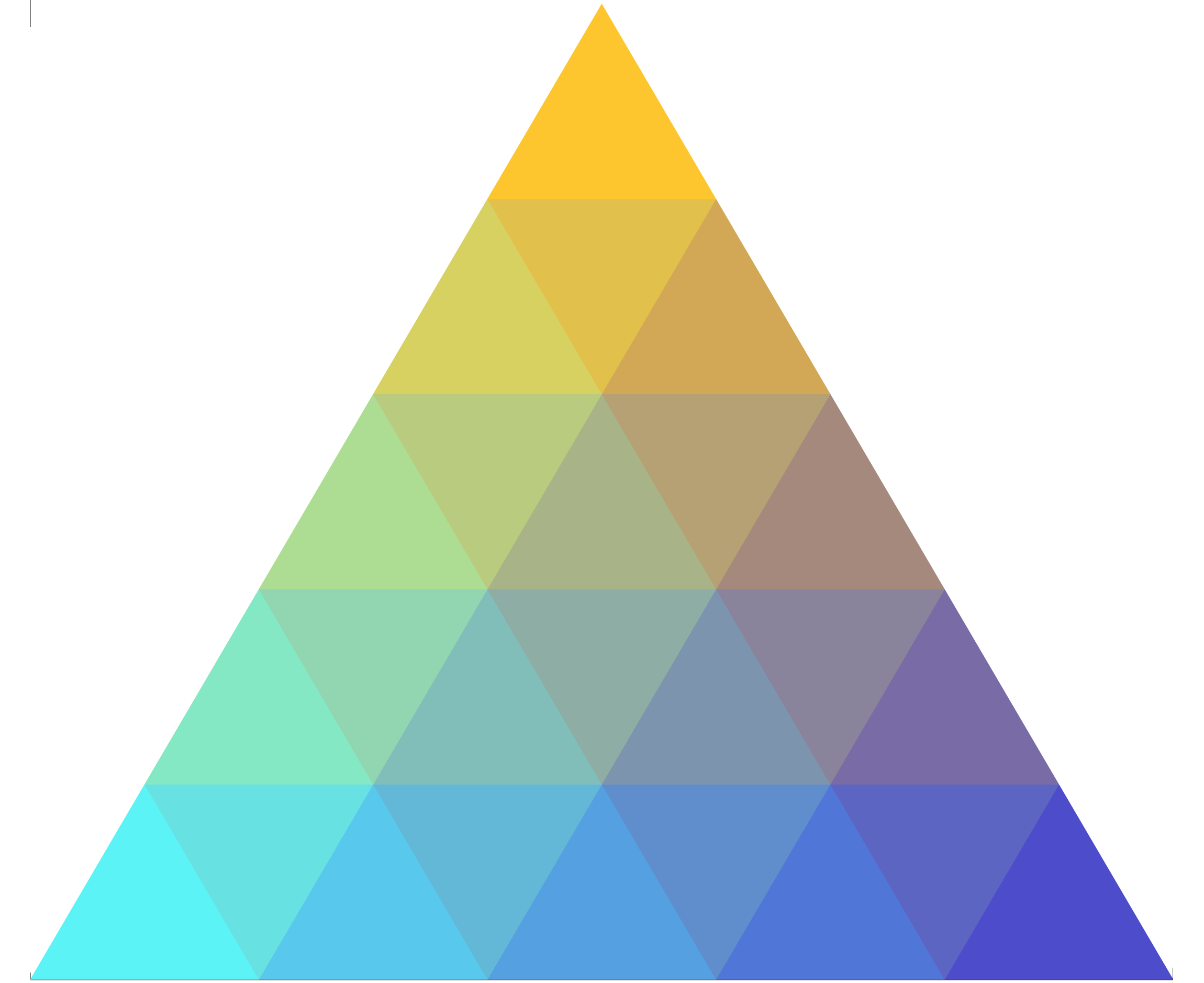}%
  }      
\subfloat[patchwork $s=10$]{%
  \includegraphics[width=0.33\columnwidth]{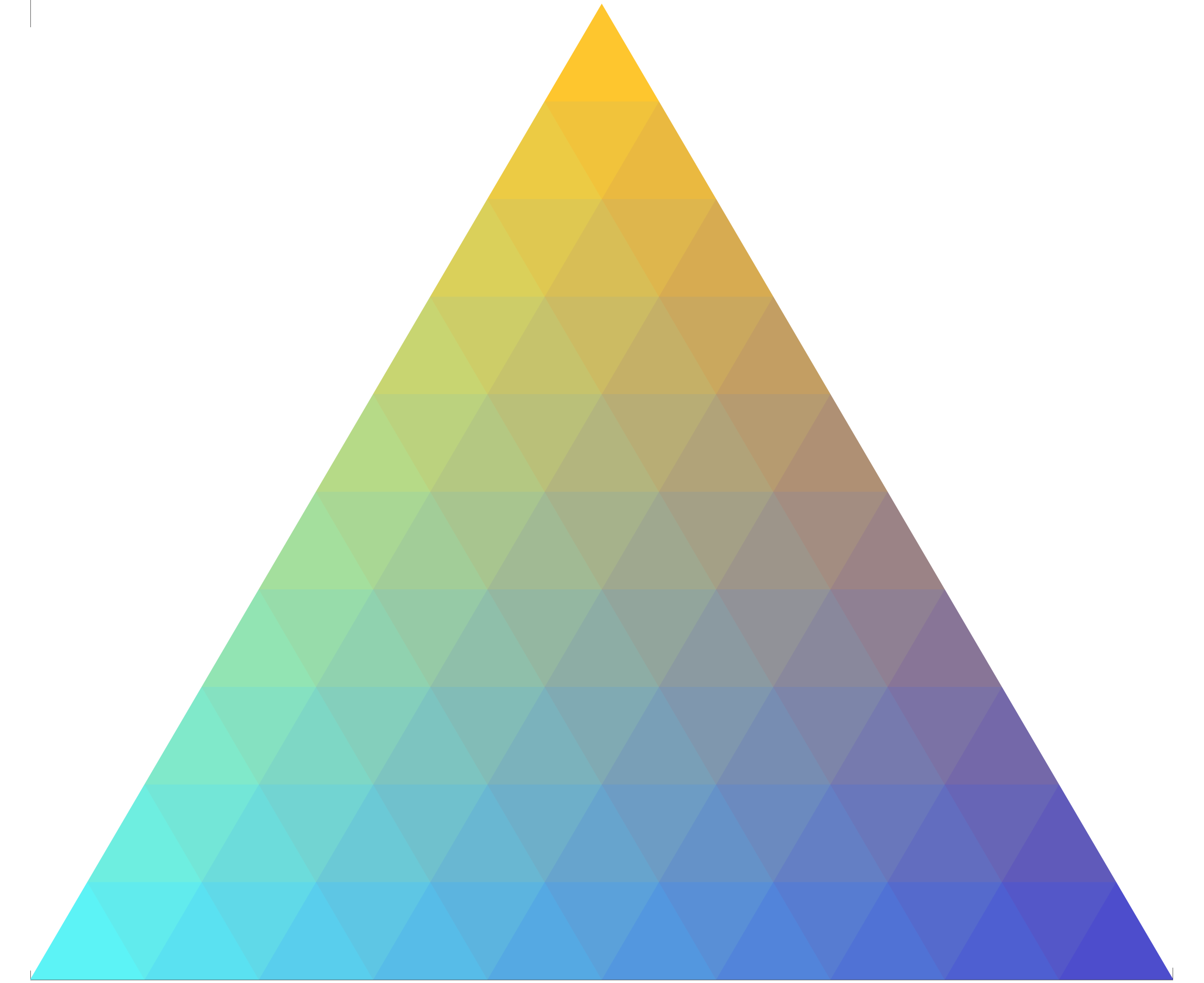}%
  }
\caption{\textbf{Discrete Interpolation:} Color sail colors for different patchwork settings $s$ (d-f) are computed using the interpolation points in (b) and (c). Interpolation with barycentric coordinates at face-centroids in (a) would undesirably exclude the three principal vertex colors and pure color gradients between any two of them.}
\label{fig:interp}
\end{figure}

\patentnote{See below that Color Sail model is not limited to RGB color space, but can be as applicable to other color spaces.}
We propose modeling any color distribution using a collection of \emph{color sails}. Each \emph{color sail} is simply a triangular patch in RGB (or other color space), defined by linear blending of its vertex colors. In addition, a \emph{color sail} has a \emph{wind} parameter, which controls the shape of the triangular patch, inflating it into a sail-like shape, and thus controlling the blending behavior smoothly (e.g.\ Fig.~\ref{fig:teaser}). Finally, each sail is equipped with a \emph{patchwork} parameter which determines how many distinct colors are present in the sail. A \emph{patchwork} setting of 0 includes only the 3 vertex colors, whereas a \emph{patchwork} setting of 15 visually approaches a continuous distribution (See Fig.~\ref{fig:interp}). 

Formally, each \emph{color sail} $\mathcal{S}$ is defined by the following parameters:
\begin{itemize}
	\item \textbf{vertex colors} $\mathbf{V} = [\bm{v}_0, \bm{v}_1, \bm{v}_2]$ : RGB colors of the 3 vertices
	\item \textbf{patchwork} $s$ : an integer parameter for the subdivision level of the triangle, taking on values 0 and above
	\item \textbf{focus point} $(p_u,p_v)$ : barycentric coordinates of the point within the triangular patch on which the wind is acting
	\item \textbf{wind} $w$ : a single float value, defining how much the sail is inflated upwind (positive) or downwind (negative)
\end{itemize}
Although this representation is compact and simple, it can model a variety of complex real-world gamuts (See Fig.~\ref{fig:patch_res}). This is not surprising, given that blended colors are common in man-made imagery, being the result of e.g. blending of paints, shading and lighting effects.

We now mathematically formulate the effect of the above parameters on the color distribution modeled by the \CS, first explaining the sail's discrete-continuous nature (\S\ref{ssec:interp}) and then the effect of wind on the blending behavior (\S\ref{ssec:wind}).

\subsection{Interpolation and Subdivision} \label{ssec:interp}

The vertex colors $\bm{v}_0, \bm{v}_1, \bm{v}_2$ define a flat triangular face in RGB space, with color at every point resulting from a smooth continuous interpolation of the vertex colors. 
However, in many graphic design applications it may be advantageous to work with discrete color distributions. For example, a graphic designer may choose to create a few discrete shading options in a flyer (See Fig.~\ref{fig:alpha_res}), rather than working with a smoothly varying gradient.

The patchwork parameter $s$ sets the subdivision level of the sail triangle, resulting in discrete gamuts for small values of $s$ and in gamuts visually approaching continuous color distributions for higher values of $s$ (Fig.~\ref{fig:interp}d-e). Let $\mathbf{C}_{\mathcal{S}} = \{\bm{c}_1 \ldots \bm{c}_n\}$ be the discrete set of colors in the distribution defined by the color sail $\mathcal{S}$. Each interpolated color can be defined as a linear combination of the vertex colors: $\bm{c}_i = u_{i0} \bm{v}_0 + u_{i1} \bm{v}_1 + u_{i2} \bm{v}_2 = \mathbf{V} \bm{u}_i$, 
where the sum $u_{i0}+u_{i1}+u_{i2}$ is constrained to be $1$ for all $i$. For a given subdivision level $s$, each $u_{ij}$ can take on values in the discrete set
$U_s = \{0,\, 1/(s-1),\, 2/(s-1),\ldots (s-1)/(s-1)\}$. Thus, we can formally define the set of colors in a color sail $\mathcal{S}$ as:
\begin{equation}\label{eq:cs0}
	\mathbf{C}_{\mathcal{S}} = \{u_{i0} \bm{v}_0 + u_{i1} \bm{v}_1 + u_{i2} \bm{v}_2\; |\; u_{ij} \in U_s,\; \Sigma_{j}\, u_{ij} = 1 \}
\end{equation}
Due to the sum constraint, the size $\|\mathbf{C}_{\mathcal{S}}\|$ is exactly $1+2+\ldots + s = s (s + 1) / 2$. \footnote{Sketch of proof: suppose $1$ is split into $(s-1)$ discrete chunks, with $s$ chunk boundaries. To pick $u_{i0},u_{i1},u_{i2}$ subject to $=1$ constraint, we must select $2$ boundaries, with and without replacement, which is exactly the sum of binomial coefficients $\binom{s}{2} + \binom{s}{1}$.} Conveniently, this is equal to the   
number of upright triangular patches in a triangle subdivided with $s$ segments per side (Fig.~\ref{fig:interp}b), and when rendering the sail, we set the colors of these triangles according to $\mathbf{C}_{\mathcal{S}}$. This results in desirable and correct behavior of including the vertex colors themselves into the set of sail colors (See interpolation points in Fig.~\ref{fig:interp}b). The unfilled upside-down triangle colors are simply based on the average of their neighbor's barycentric coordinates $\bm{u}_i$ (Fig.~\ref{fig:interp}c) and we expand $\mathbf{C}_{\mathcal{S}}$ to include them. Conceptually this is equivalent to shrinking the triangle vertices by a factor of $(s-1)/(s*\sqrt{3})$, to align with the centroids of the subdivided triangles. We do not use the barycentric coordinates for the centroids themselves, as this would exclude the three vertex colors themselves and gradients involving only two of them.

\begin{figure}[ht]
\centering
\subfloat[Bezier triangle control points; effect of varying $(p_u,p_v)$ for $w=0$]{%
  \includegraphics[width=0.95\columnwidth]{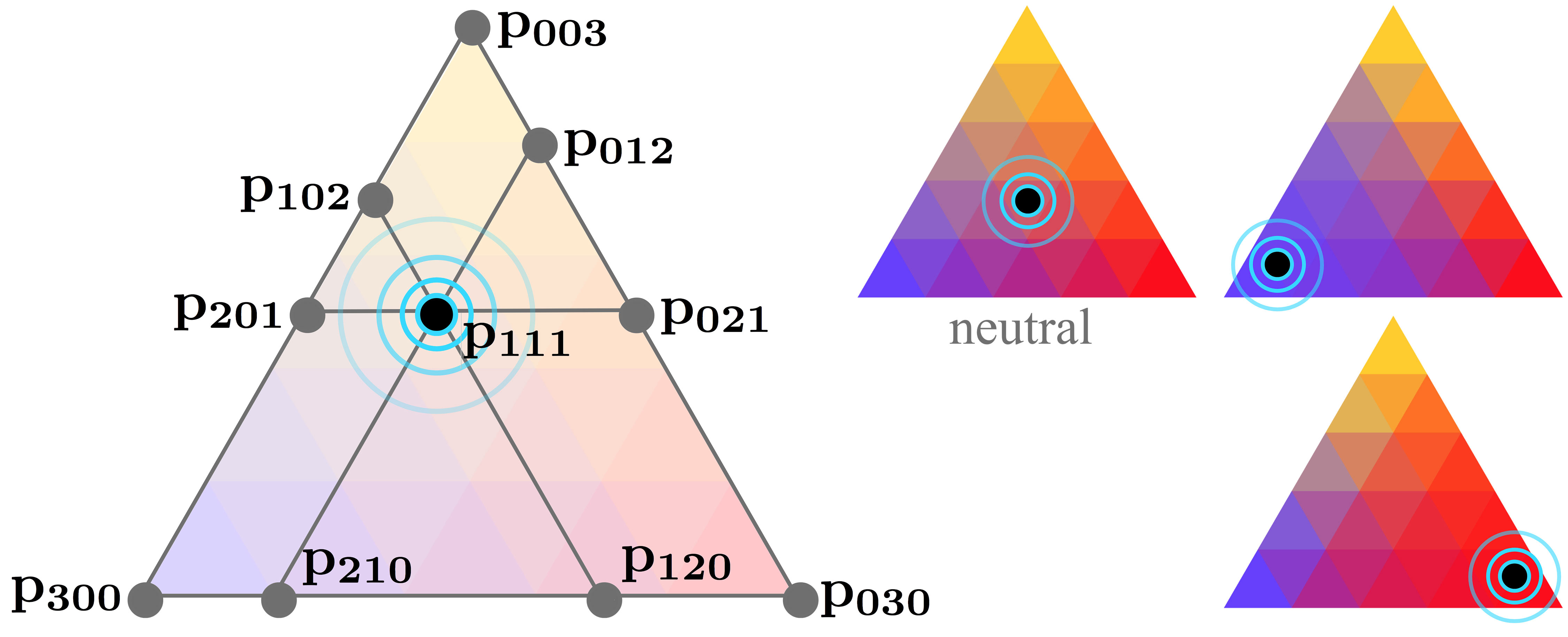}%
  }
\caption{\textbf{Wind:} In the presence of wind, color sail colors are defined using a Bezier triangle with two interactive controls: the strength of the wind $w$, and the location of the focus point $\bm{p}_{111}$, which has the effect of brining attention to a particular vertex (see right). }
\label{fig:wind}
\end{figure}

\subsection{Wind and Blending} \label{ssec:wind}

Simple linear interpolation in RGB can be quite limiting, as it is well-known that many natural media exhibit non-linear color blending behavior (Fig.~\ref{fig:km}a). For this reason, paint blending and compositing is often modeled using a variety of alternative approaches (See \S\ref{sec:prev}, Color blending). Rather than modeling a particular medium, we define color sail \emph{wind} parameters that allow the sail to model an array of smooth blending behaviors (Fig.~\ref{fig:km}). 

To model non-linear blending behavior, we use a cubic Bezier triangle \cite{farin2002curves}, and allow interactive control of its shape using two parameters: the barycentric coordinates of the focus point $(p_u,p_v)$ on which the wind is acting and the wind strength parameter $w$. The shape of the Bezier triangle is defined by $10$ control points $\mathbf{P} = \{\bm{p}_{ijk}, i+j+k=3\}$ in RGB (Fig.~\ref{fig:wind}a). Specifically, the 3D location $\bm{c}_{\mathbf{P}}$ of a planar color sail point parameterized by barycentric coordinates $(u_{0}, u_{1}, 1-u_{0}-u_{1})$ over the triangle vertices (such as our planar interpolated colors in Eq.~\ref{eq:cs0}), can now be expressed in terms of $\bm{p}_{ijk}$ using Bernstein polynomials $\mathit{B}_{ijk}$:
\begin{align}
\bm{c}_{\mathbf{P}}(u_{0}, u_{1}) &= \sum_{i+j+k=3} \mathit{B}_{ijk}(u_0,u_1)\bm{p}_{ijk}\\
 \mathit{B}_{ijk}(u_0,u_1) &= \frac{3!}{i!j!k!}u_0^iu_1^j(1-u_0-u_1)^k
\end{align}
To allow simple interactive control, we define the location of the sail control points $P_{\mathcal{S}}$ in terms of the focus point $(p_u,p_v)$ and wind $w$, as well as the vertex colors $\mathbf{V}$: 
\begin{equation}\label{eq:pts}
\bm{p}_{ijk} = \mathbf{V} \bm{u}_{ijk}(p_u, p_v) + f(d^2_{ijk}) \cdot w \vec{n}
\end{equation}
where the first term is the point location in the triangle plane, and the second term is its displacement in the normal direction. First, each control point $\bm{p}_{ijk}$ is assigned barycentric coordinates $\bm{u}_{ijk}$ with respect to the triangle vertices $\textbf{V}$. We define $\bm{u}_{111} = (p_u,p_v,1-p_u-p_v)$ as the focus point, and set corner control points $\bm{u}_{300}$, $\bm{u}_{030}$, $\bm{u}_{003}$ to the vertex colors. The remaining coordinates are likewise naturally expressed in terms of $(p_u,p_v)$, as shown in Fig.~\ref{fig:wind}a. Then, we displace all control points $\bm{p}_{ijk}$ except the corner points in the direction $\vec{n}$ normal to the triangle plane. The focus point control point $\bm{p}_{111}$ is displaced the most, and the displacement of other control points falls off with the distance squared $d^2_{ijk}$ to the central control point. (See Appendix~\ref{app:interp} for details.) 

 Thus, the set of non-linearly interpolated colors defined by the color sail $\mathcal{S}$ is:
\begin{equation}\label{eq:cs1}
	\mathbf{C}^w_{\mathcal{S}} =
	\{ \bm{c}_{\mathbf{P}}(u_{i0}, u_{i1})\; |\; u_{ij} \in U_s,\; \Sigma_{j}\, u_{ij} \leq 1,\; \mathbf{P} = \mathbf{P}_{\mathcal{S}} \}
\end{equation}
where $u_{i0}$, $u_{i1}$ are exactly as in Eq.~\ref{eq:cs0}, and the 10 control points $\mathbf{P}$ interpolated by $\bm{c}_{\mathbf{P}}$ are expressed in terms of the color sail parameters. Our specific choices (See ~\ref{app:interp}) make it possible to express $\mathbf{C}^w_{\mathcal{S}}$ in matrix form with clean derivatives of the interpolated colors with respect to the color sail parameters.

\subsection{Interpretation and Prior Art}

In a continuous setting, color sail is a generalization of the  two-color gradient, a ubiquitous tool in graphic design, to 3 dimensions. Just as the gradient tool allows creating an intermediate control point to sway the interpolation from its linear path, the wind $w$ allows interpolated colors to deviate smoothly from planar blending. Just as the middle control point can be shifted within the 2D gradient to give more space to the ranges of colors on either side, bringing the focus point close to a vertex biases the interpolation toward that color. This allows the artist to narrow in on fine variations of a particular color (Fig.~\ref{fig:wind}, right). Unlike other 3D color manifold representations~\cite{nguyen2015data}, ours is sparse and allows interactive control. We note that the layout of the color sail on the screen is not fixed: the artist can deform the triangle in any way for easier exploration and color picking. Color sail has a fixed number of parameters, unlike other interactive continuous gamuts ~\cite{shugrina2017playful}, making it more amenable to machine learning. The discrete-continuous setting gives color sail an additional degree of flexibility, varying the number of represented colors with no change to the required storage. 
 
\edits{While we do not explore extensions of the \CS{}, such as linking of colors between different color sails, allowing global adjustments of colors with lightness and saturation controls, or support of degenerate 2 and 1 color color sails (lines and points), such extensions would be easy to prototype with minor adjustments to the representation we present here. In addition, we believe that exciting interactive opportunities exist for the control of the wind parameter, for example, by learning wind behavior compatible with the blending properties of particular media. }

\section{Color Sail Rig}\label{sec:rig}
\begin{figure}[ht]
\includegraphics[width=0.99\columnwidth]{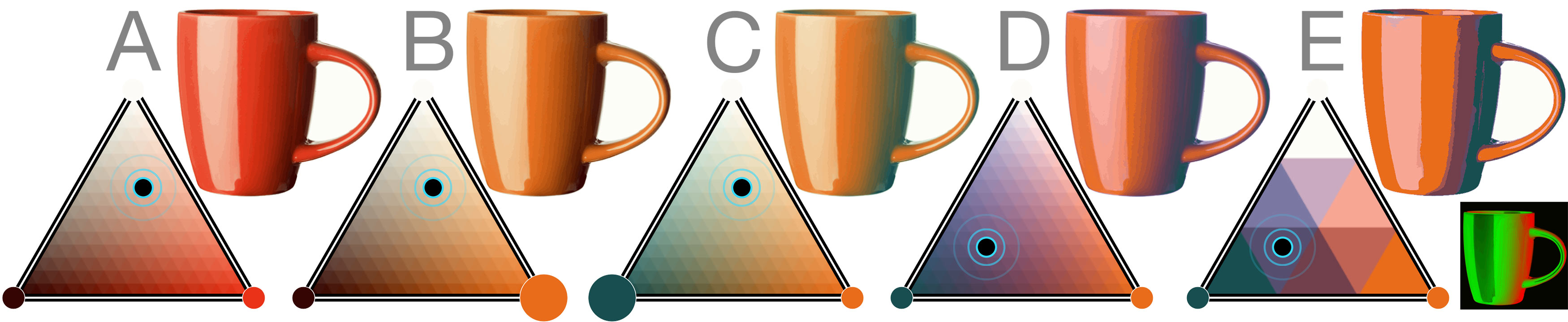}%
\caption{\textbf{Editing with a Color Sail Rig:} The mapping from an image to a color sail (bottom right) allows an artist to explore variations of base colors (A,B,C), as well as color focus and wind blending behavior (D), and discreteness (E).
}
\label{fig:rig}
\end{figure}

An interactive color sail can be a useful tool to explore color gamuts independently of any artwork, but its 
real power comes to light when it is linked to an image. We propose modeling a color distribution of an artwork or design in a piecewise manner using a collection of color sails. It is possible to change colors of an artwork in our model using a mapping from an image to its color sails, akin to establishing a mapping from a discrete color palette to colors in a graphic design to allow for palette based recoloring. This allows interactive exploration of an artwork's colors by changing vertex colors (Fig.~\ref{fig:rig} B,C), focus point and wind (Fig.~\ref{fig:rig} D), and patchwork level (Fig.~\ref{fig:rig} E) of its color sails, resulting in a wide range of visual effects. We refer to a collection of color sails and the mapping from image pixels to its color sails as a \emph{color sail rig}. While it is possible to manually establish a mapping between the palette and the artwork during art authoring~\cite{shugrina2017playful}, it is unrealistic to expect artists to use a single palette UI for their work. \edits{We therefore aim to construct a color sail rig automatically (\S\ref{sec:riglearn}) in order to drive interactive exploration (\S\ref{sec:ux}).}

\section{Learning Color Sail Rigs}\label{sec:riglearn}
\patentnote{We are working on developing a better method for point 3 below, the mapping of pixels to a color in the color sail.}
The problem of rigging an image with sails has three subparts: 1) separating an image into regions well modeled by color sails, 2) fitting a color sail to every region, 3) establishing a mapping between each region's pixels and its color sail. 
In our work, we first address problem 2 (\S\ref{sec:pg}), and then expand scope to 1 (\S\ref{sec:learning}). \edits{We largely leave problem 3 to future work, and use a simple approach (\S\ref{ssec:mapping}).}

\begin{figure*}[ht]
\centering
\subfloat[Full Alpha Network architecture uses the Palette Network.]{%
  \includegraphics[width=0.46\linewidth]{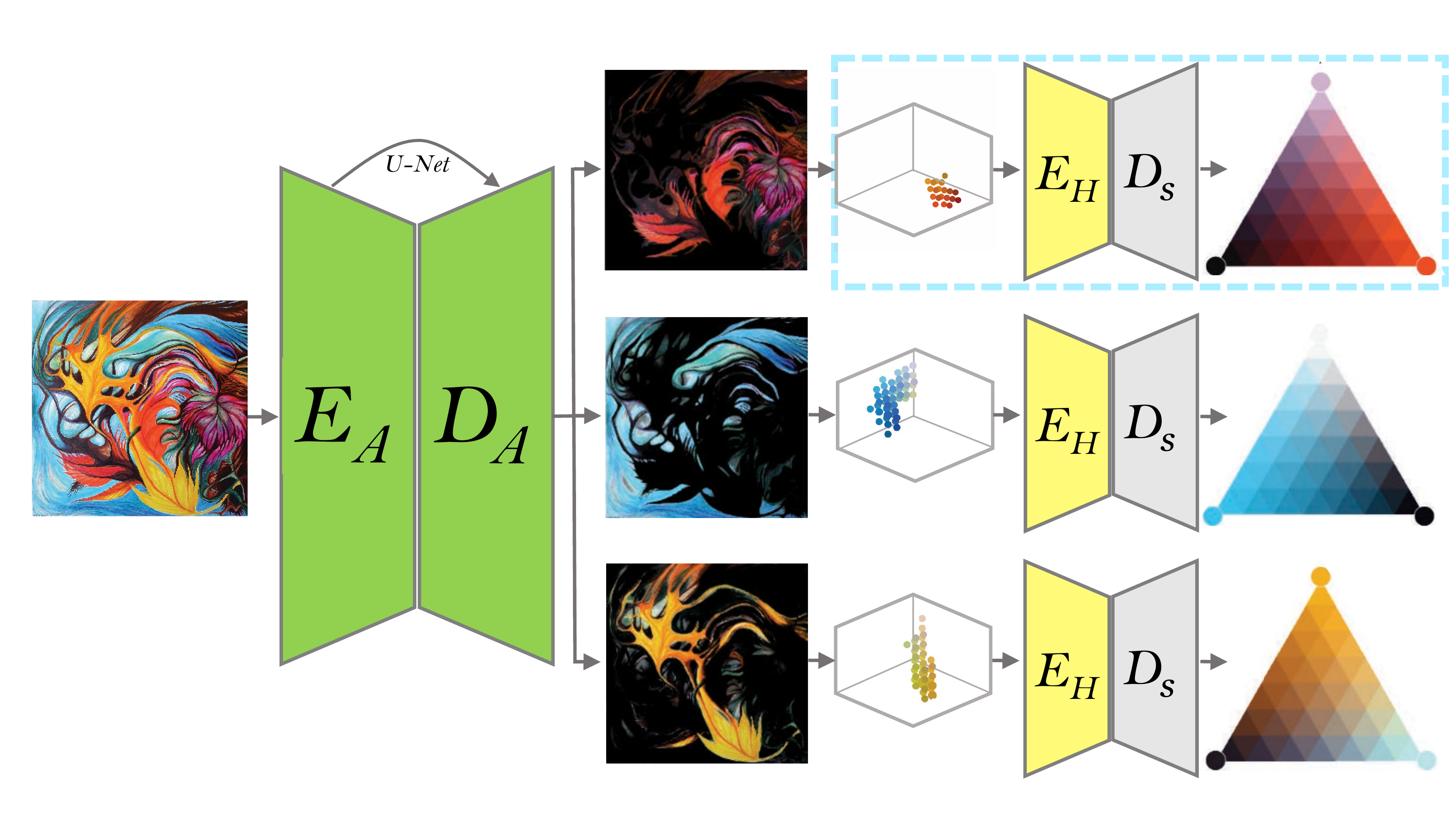}%
  }
  \subfloat[Zoomed in view of the Palette Network.]{
  \includegraphics[width=0.455\linewidth]{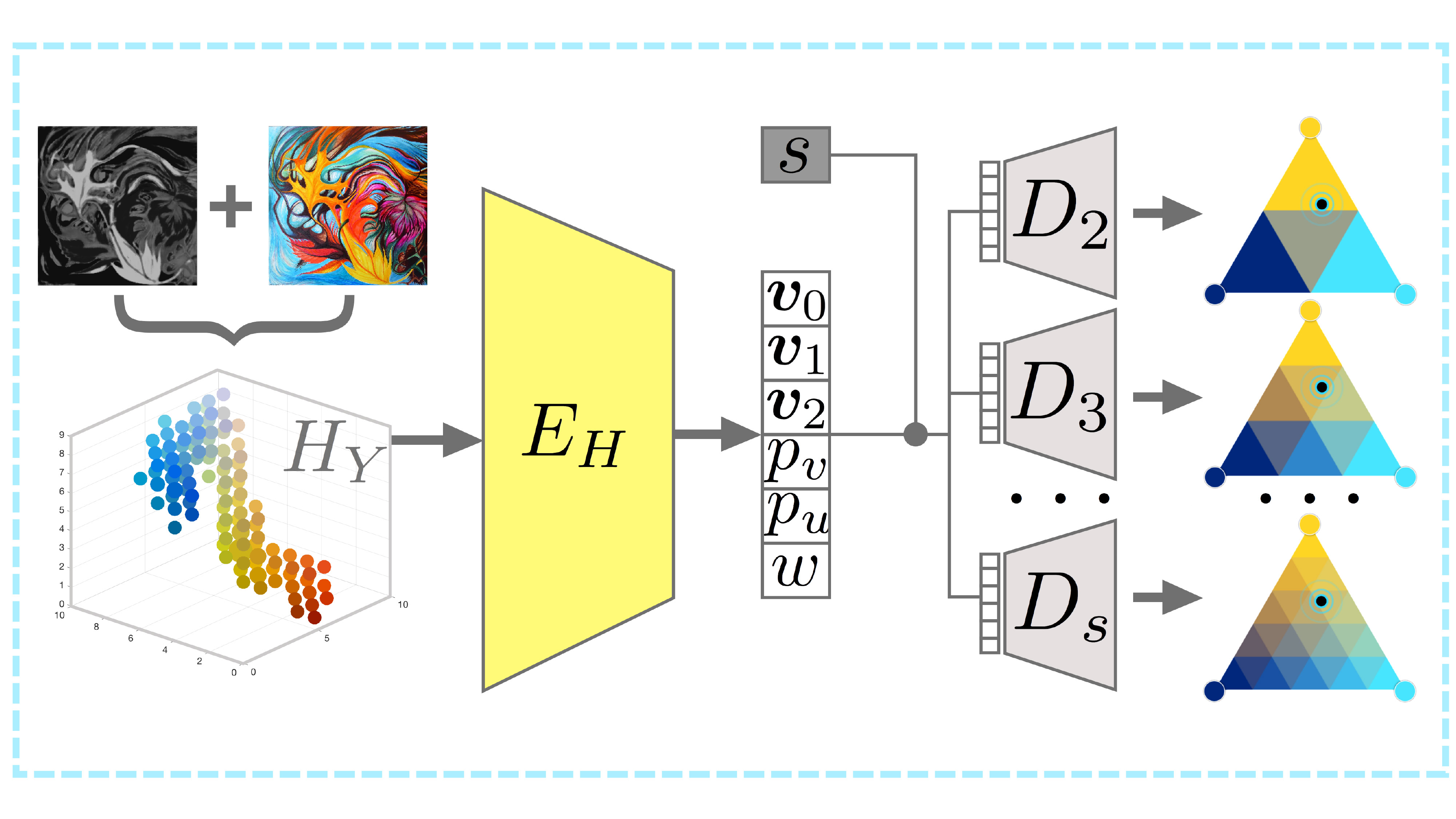}%
  }
\caption{\textbf{Full Architecture:} a) Alpha Network's U-Net architecture converts an input image into $N_{\alpha}$ alpha masks (\S\ref{sec:learning}), which are used to weigh histograms passed to the Palette Network (\S\ref{sec:pg}) to produce a color sail per alpha mask. b) Palette Network (zoomed view) encodes a normalized histogram input as the \emph{color sail} parameters, "decoded" into the full set of color sail colors $\mathbf{C}^w_{\mathcal{S}}$ using a deterministic decoder $D_s$, constructed for a particular patchwork value $s$.}
\label{fig:full-net}
\end{figure*}

%

Fig.~\ref{fig:full-net}a demonstrates our full model architecture. Given an image $\mathbf{Y}$, we produce $N_\alpha$ alpha masks (\S\ref{sec:learning}). For each alpha mask, we encode the corresponding colors into a histogram and output a single color sail (\S\ref{sec:pg}). The model is trained to be able to predict alpha masks such that the color distribution in the region under the mask can be explained using a single color sail. While the model runs end-to-end, we train the palette prediction network separately at first. Our choice is motivated by two reasons: 1) color sail fitting is an independent problem, a solution to which may be useful outside of the context of \st{semantic color} \edits{image} segmentation, 2) a pre-trained palette graph allows the \emph{alpha network} to focus on learning \st{semantics} \edits{segmentation} without conflating its search direction with a separate simple task. We lay out the details of the palette and alpha mask networks in the following subsections.

\subsection{Learning Palette Encoding}\label{sec:pg}

For any input set of colors $\mathbf{Y}$, our goal is to fit the parameters of a color sail $\mathcal{S}$ in order to minimize some error metric $\mathcal{L}(\mathbf{Y}, \mathbf{C}^w_{\mathcal{S}})$ between the input colors and the set of color sail colors $\mathbf{C}^w_{\mathcal{S}}$ (\S\ref{sec:rep}). In particular, the goal is to find 3 color vertices $\bm{v}_0; \bm{v}_1; \bm{v}_2$, patchwork $s$, pressure point $(p_u, p_v)$ and wind $w$ to best approximate $\mathbf{Y}$. Although a number of iterative optimization approaches are possible to accomplish this task, training a neural network is a natural solution, given the differentiability requirement for this module to function inside our alpha mask neural network (\S\ref{sec:learning}).


 While it may be tempting to work with image input directly, no spatial knowledge is necessary: our goal is simply to approximate a set of colors as closely as possible by a deformed planar patch of RGB space. Our method must be able to approximate colors in regions of varying size and respect alpha weights assigned to pixels during region learning (\S\ref{sec:learning}). For these reasons, we chose to represent the input color distribution as a histogram $H_Y$ of $n\times n\times n$ bins in RGB, where the count of a pixel's color can be weighted by its alpha and the bin values are normalized. This means that the input size is always $n^3$, regardless of the region size (we use $n=10$).
 
 \subsubsection{Palette Network} Our network has an encoder-decoder structure, with the encoder $E_H$ mapping the input histogram $H_Y$ to the parameters of a color sail $\mathcal{S}$ (Fig.~\ref{fig:full-net}b). Unlike a typical decoder setup, palette network's decoder is fully deterministic, as it simply applies the formula for computing the interpolated colors $C^w_{\mathcal{S}}$ based on the color sail parameters (Eq.~\ref{eq:cs1}). We choose to avoid the complexity of encoding the integer patchwork parameter $s$, which controls the number of output interpolated colors. Instead, $s$ is an external input, which acts as a switch, determining which decoder $D_s$ is currently active. This can be a powerful parameter both during network training and at run time. We found that a simple 4-layer fully connected network performs well on this problem, and that the determining factors of success are the choice of the loss function and data selection (\S\ref{ssec:pg_eval}). \patentnote{In the improved version, $s$ will also be learned automatically for any given distribution.} 

\subsubsection{Evaluation Criteria}

Although the input to the palette network is a histogram, it is more natural to define error metrics on the pixel colors $\bm{y}_i$ in the input image $Y$. First, a representative palette must contain all the colors in $Y$. We track this criterion with two metrics. First, we compute a greedy $L_2$ RGB reconstruction loss: 
\begin{equation}\label{eq:e_l2}
	E_{L2}(Y, \mathbf{C}^w_{\mathcal{S}}) = \frac{1}{\|Y\|} \sum_{\bm{y}_i \in Y} \; \underset{\bm{c}_j \in \mathbf{C}^w_{\mathcal{S}}}{\operatorname{min}} \; \| \bm{y}_i - \bm{c}_j \|_2
\end{equation}
This metric tracks the average error for a pixel in the image if it were modeled using the best matching interpolated color in $\mathbf{C}^w_{\mathcal{S}}$. However, $L2$ loss does not necessarily correlate well with perceptual metrics due to averaging effects. In addition, we track the fraction of the patch pixels that are well-approximated by the color sail colors:
\begin{align}\label{eq:e_perc}
	R_{\%}(Y, \mathbf{C}^w_{\mathcal{S}}) &= \frac{1}{\|Y\|}\sum_{\bm{y}_i \in Y} \mathds{1}\, \big[\; \underset{\bm{c}_j \in \mathbf{C}^w_{\mathcal{S}}}{\operatorname{min}} \; \|lab(\bm{y}_i) - lab(\bm{c}_j)\| < \delta \;\big]\\
	E_{\%}(Y, \mathbf{C}^w_{\mathcal{S}}) &= 1 - R_{\%}(Y, \mathbf{C}^w_{\mathcal{S}})
\end{align}
where $\mathds{1}$ is $1$ if the condition inside the brackets evaluates to true and $0$ otherwise. Essentially, we compute the fraction of pixels that have a "good enough" approximation using the best matching color in the sail color. It is possible to define a consistent "barely noticeable" distance in CieLAB space, where distances are more perceptually uniform (we use $\delta=10$, see Fig.~\ref{fig:patch_res}). The loss version of this is then $E_{\%}$, i.e.\ the fraction of pixels \emph{not} well represented. 

A good palette must not only be able to reconstruct the image colors, but also contain few irrelevant colors. For instance, a red patch $Y$ will have a zero $E_{L2}$ and $E_{\%}$ losses if one of the vertex colors $\bm{v}_i$ is red, even if the other vertices have colors that bear no connection to $Y$. There is no clear way to measure how \emph{relevant} a palette color is to an image.\footnote{E.g.~inverse reconstruction loss is meaningless here, as a single matching pixel in the image would render a palette color "relevant", which is clearly not the case, especially if $Y$ has some noise or compression artifacts.} We choose a measure that encourages the distribution of colors in the output color sail $\mathcal{S}$ to resemble the color distribution of $Y$, namely Kullback-Leibler divergence:
\begin{equation}\label{eq:e_kl}
E_{KL} = KL(H_{\mathcal{S}} || H_{Y}) = - \sum_{b \in H_{\mathcal{S}}} H_{\mathcal{S}}(b) \operatorname{log} \frac{H_Y(b)}{H_{\mathcal{S}}(b)}
\end{equation}
where $H_{\mathcal{S}}$ is the normalized histogram computed over the colors $\mathbf{C}^w_{\mathcal{S}}$, with equal count assigned to every color. The asymmetric nature of this measure is a good fit for the question about palette colors in particular, but it is not perfect: a palette need not contain colors in the same proportion as the image itself. To mediate this discrepancy, we compute $H_Y(b)$ for Eq.~\ref{eq:e_kl} by fist computing small patch histograms, then taking the max of every bin across patches, and normalizing. This encourages every significant color in $Y$ to attain a maximal value in the histogram.

\subsubsection{Learning} 

Color sails are designed to represent color distributions of coherent image regions, where a single low-order manifold in RGB space provides enough representative power. For this reason, we do not aim to fit color sails to entire images. Instead, the input images $\mathbf{Y}$ in our train and test data are random image patches. This data provide an approximation for coherent regions the colors of which we would like to model. Our explicit aim is to support visual creative tasks, such as digital painting, graphic design, and visualization, and we select our train and test sets accordingly (\S\ref{ssec:datasets}).

While $E_{\%}$ is a loss we would most like to optimize, it is not amenable to gradient descent based optimization techniques. Therefore, for our loss function $\mathcal{L}(\mathbf{Y}, \mathbf{C}^w_{\mathcal{S}})$ we use a weighted sum of $E_{L2}$ and $E_{KL}$ (See \S\ref{ssec:pg_eval}). We train the weights of our encoder $E_H$ using stochastic gradient descent with the Adam optimizer ~\cite{kingma2014adam} (learning rate $10^{-3}$) by back propagation through the entire network, including the deterministic palette decoder $D_s$, which itself has no trainable weights. 

\subsubsection{Runtime}\label{ssec:pg_interact}
\edits{After the palette graph has been trained, we only use the encoder part $E_H$ at runtime. Given any input histogram $H$, e.g.\ computed for an image or a soft region, $E_H$ can encode it into the parameters of a color sail $\mathcal{S}$. While this already facilitates a variety of interactive possibilities~\S\ref{sec:ux}, a single color sail lacks representative power to model many distinctly colored regions of the image, motivating the next section.}

\begin{figure*}[ht!]
  \centering
  \includegraphics[width=7.2in]{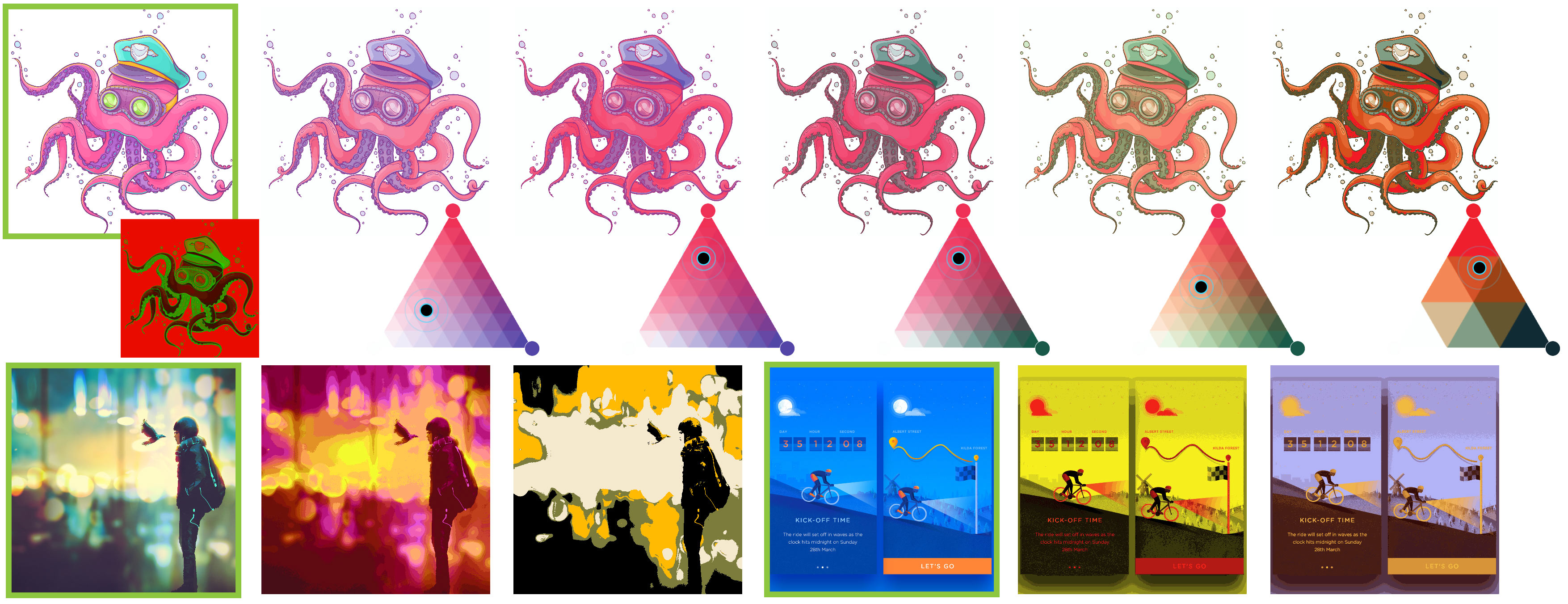}
  \vspace{-5mm}
  \caption{\textbf{Single Color Sail Editing:} Results of editing designs with a single automatically generated color sail. Top row: original (uv color sail mapping as inset), best reconstruction using a single color sail, user edits. Bottom row: more examples of color sail exploration.}
  \label{fig:one_sail_edit}
\end{figure*}

\subsection{Alpha Mask Learning}\label{sec:learning}

\edits{In order to model a wider array of disparate color gamuts, we propose to automatically segment image into regions, each governed by its own color sail. This would allow targeted region-based color editing using the obtained color sail rig. }Specifically, for an input image $\mathbf{Y}$, the goal is to predict alpha masks and a color sail for each alpha mask such that all colors in $\mathbf{Y}$ are well represented in the complete color sail rig. We emphasize that the alpha masks are learned without any explicit supervision. We leverage our trained palette network (\S\ref{sec:pg}) and optimize an image reconstruction loss to learn meaningful alpha masks. Optimizing for image reconstruction given a constant palette network, forces the alpha network to output masks that correspond to image regions that can be well explained by a distribution of colors in a single color sail. We detail our model architecture and learning techniques below.

\subsubsection{Alpha Network}\label{sec:alpha_net}
Following the recent trend in Image Segmentation literature, we use an encoder decoder architecture~\cite{long2015fully} for our alpha network. Additionally, we add skip connections from encoder layers to decoder layers, akin to the U-Net architecture~\cite{ronneberger2015u}. The encoder-decoder network $E_AE_D$ (Fig.\ref{fig:full-net}) takes as input an RGB image and outputs $N_\alpha$ channels at the input resolution.
To produce the final alpha masks we apply a softmax function across channels, thus ensuring that the sum of alphas at every pixel is $1$.
We construct histograms $H_i$ of RGB colors as in \S\ref{sec:pg} per alpha mask $A_i$ by adding soft votes (i.e. the alpha values) into corresponding bins. Each (normalized) histogram is then passed through our palette network $E_H$ to produce the parameters of a single color sail $\mathcal{S}_i$, which are then decoded into its full set of colors $\mathbf{C}_{\mathcal{S}_i}$ using the decoder $D_H$ (\S\ref{sec:pg}).

\subsubsection{Learning}
We train the network to be able to reconstruct the original image using the produced alphas and color sails. Let us denote the $i^{th}$ alpha mask as $A_i$ and the corresponding set of color sail colors as $\mathbf{C}_{\mathcal{S}_i}$ (Eq.~\ref{eq:cs1}). Using these, we compute the pixel value at location $x,y$ of the reconstructed image $\mathbf{Y}_R$ as:
\begin{equation}
\label{eq:recon}
\mathbf{Y}_R(x,y) = \sum_{1 \leq i \leq N_\alpha}A_i(x,y)\;  \underset{\bm{c}_j \in \mathbf{C}_{S_i}}{\arg\min} \|Y(x,y) - \bm{c}_j\|_2
\end{equation}
Our reconstruction loss is simply the mean $L2$ loss between the pixels of $\mathbf{Y}$ and $\mathbf{Y}_R$, similarly to the palette learning (Eq.~\ref{eq:e_l2}). To promote spatial regularization in the alphas, we add total variation regularization on the predicted alphas to the optimization objective (with a weight of $10^{-3}$). In addition, to encourage the network to select binary masks, rather than blend colors, we introduce a temperature parameter $\tau$ in our softmax:
\begin{equation}
\operatorname{softmax}(\pmb{z},\tau)_j = \frac{e^{z_j/\tau}}{\sum\limits_{k=1}^{K}{e^{z_k/\tau}}}
\end{equation}
Lower values of $\tau$ make the predictions more confident, and empirically we found $\tau=\frac{1}{3}$ to provide a good tradeoff between allowing blending and preferring binary alphas. To train the model, we used the Adam optimizer~\cite{kingma2014adam} with a learning rate of $10^{-3}$ and other default parameters.
We also note that while training the alpha network, we initialize the palette network from a pre-trained checkpoint and keep its weights frozen. More details about the network structure can be found in the Appendix.


\subsubsection{\edits{Runtime}}
Different images are best represented with different numbers of color sails. We do not explicitly address this concern in this work, but instead train multiple models for producing different number of alphas $N_\alpha$ per image (specifically, $N_\alpha$=2,3,4,5). At runtime, we select the model for an image $\mathbf{Y}$ using the following equation:
\begin{equation}
N_\alpha(\mathbf{Y}) = \underset{N_\alpha}{\arg\min}(L_{N_\alpha}(\mathbf{Y}) + 100N_\alpha)
\label{eq:n_alpha}
\end{equation}
where $L_{N_\alpha}(\mathbf{Y})$ is the weighted sum of our reconstruction and regularization losses. We arrived at this trade off after inspecting results on our validation dataset with the aim of producing the minimum number of alphas while making sure that the color sail rig represents the image well enough. 



\subsection{Pixel Mapping}\label{ssec:mapping}
\edits{
We can pass an image histogram through our palette encoder $E_H$ to obtain one representative color sail, or use the full alpha network (Fig.~\ref{fig:full-net}a) to obtain image alpha masks $A_i$ and corresponding color sails $\mathcal{S}_i$ for images with more varied gamuts. Once the color sail parameters have been computed, we must establish a relationship between the colors $\mathbf{C}_{\mathcal{S}_i}$ in each color sail and pixels in order to drive image recoloring by edits to each color sail. We largely leave this problem to future work, and simply map each pixel to the closest color in each color sail. After the user edits to any linked color sail, we reconstruct recolored pixel value following Eq.~\ref{eq:recon}, where $argmin$ is over the original color values in the image and the color sail. Our results show that even this simple mapping (Fig.~\ref{fig:rig} bottom right, visualized as $u$-$v$ coordinates in the sail) can produce powerful color exploration results. However, there is much room for improvement in linking image colors to color sails, perhaps respecting semantics or smoothness constraints, or even adjusting based on edit type.}


\section{Using Automatic Rigs}\label{sec:ux}
\edits{Once an image is automatically rigged with alpha masks $A_i$, corresponding color sails $\mathcal{S}_i$, and a pixel to color sail mapping, the user can interactively edit the parameters of each color sail in order to recolor the artwork. At runtime, we do not use the palette decoder $D_s$ (\S\ref{sec:pg}) that computes color sail colors $\mathbf{C}_{\mathcal{S}_i}$ based on color sail parameters. Our web-based user interface implements its own version of $D_s$, able to quickly provide updates to the rendering pipeline during user interaction. As the user is manipulating the color sail parameters, the interactive decoder $D_s$ renders the updated sail, and every linked pixel is recolored with the linked color(s) according to Eq.~\ref{eq:recon} (with $argmin$ computed once over the original colors, as in \S\ref{ssec:mapping}).}

\edits{Even in the case of a single color sail fit to the entire image, color sail manipulation opens up a wide range of color editing effects, while respecting color relationships in the image. For example, by changing vertex colors, the shape of the gradient in RGB space, and the discreteness level of the sail an artist can explore many possibilities for the design (Fig.~\ref{fig:one_sail_edit}). In the case of multiple alpha masks $A_i$ and color sails $\mathcal{S}_i$, the artist can perform more targeted edits by choosing which color sail to work with. We provide a visualization of the affected soft region in the interface. This allows fast exploration of widely different color combination for various regions of the image (See \S\ref{ssec:alpha_eval}, \S\ref{sec:user_study}). Demos are available at \textbf{\href{http://www.colorsails.com}{www.colorsails.com}}. }

\edits{Given a new image, only one forward pass in the network is required to compute the full color sail rig (170 milliseconds on nVidia GTX Titan X). This makes it possible for artists to interactively correct automatic alpha masks, or to provide their own. In this case, color sail fitting to existing alpha masks would take a fraction of this time. Recoloring process itself is done interactively using our web interface written in Javascript without any special optimizations (every pixel is recolored during every interaction). For larger image resolution, or larger patchwork $s$, a number of straightforward optimizations would easily keep the entire process interactive.  }




\section{Results}\label{sec:results}

\subsection{Palette Encoding Results}\label{ssec:pg_eval}
\begin{figure}[ht]
\centering
\subfloat[$E_{L2}$ plot, unnormalized for 32x32]{%
  \includegraphics[width=0.47\columnwidth]{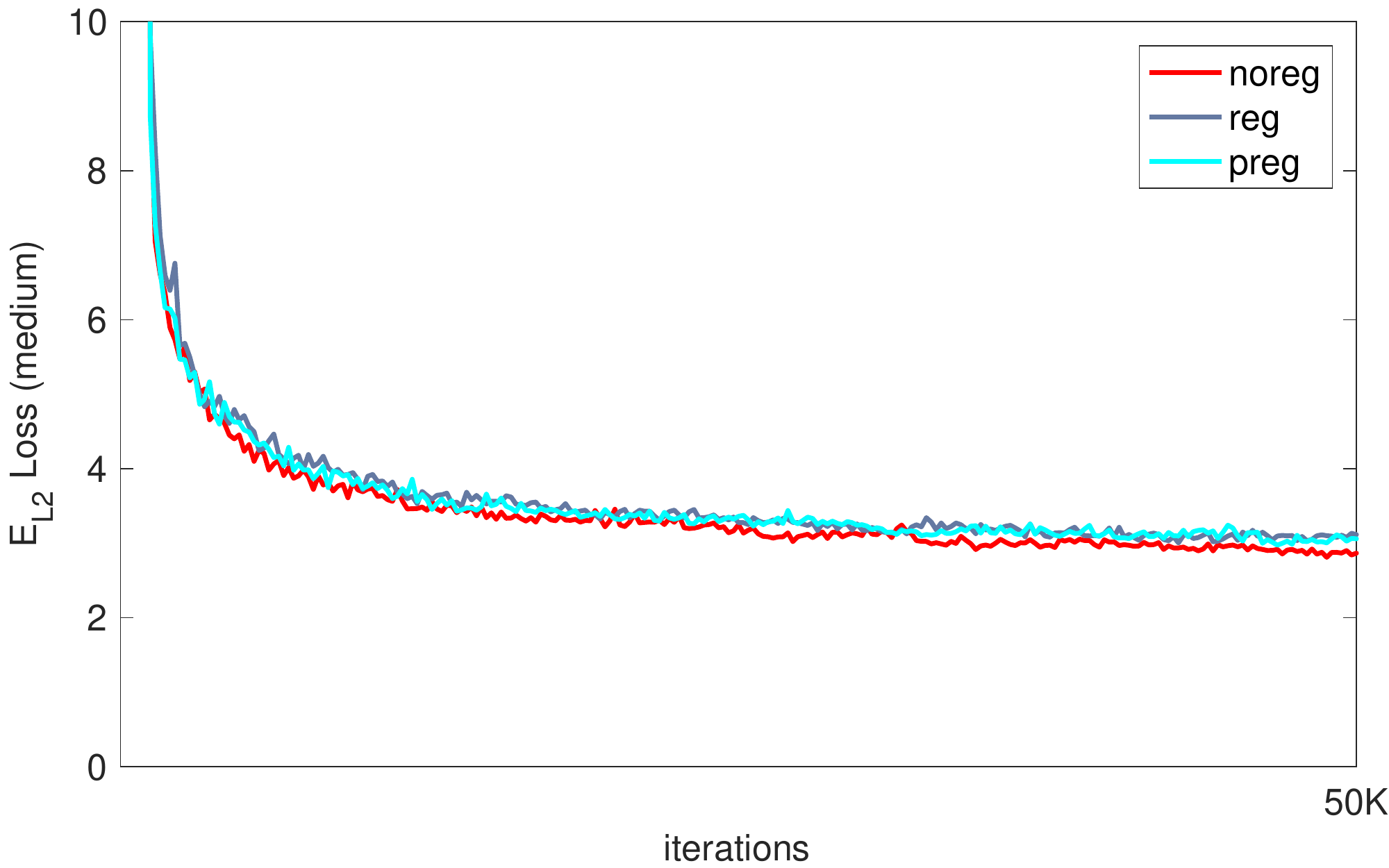}%
  }
\subfloat[$E_{KL}$ plot]{%
  \includegraphics[width=0.47\columnwidth]{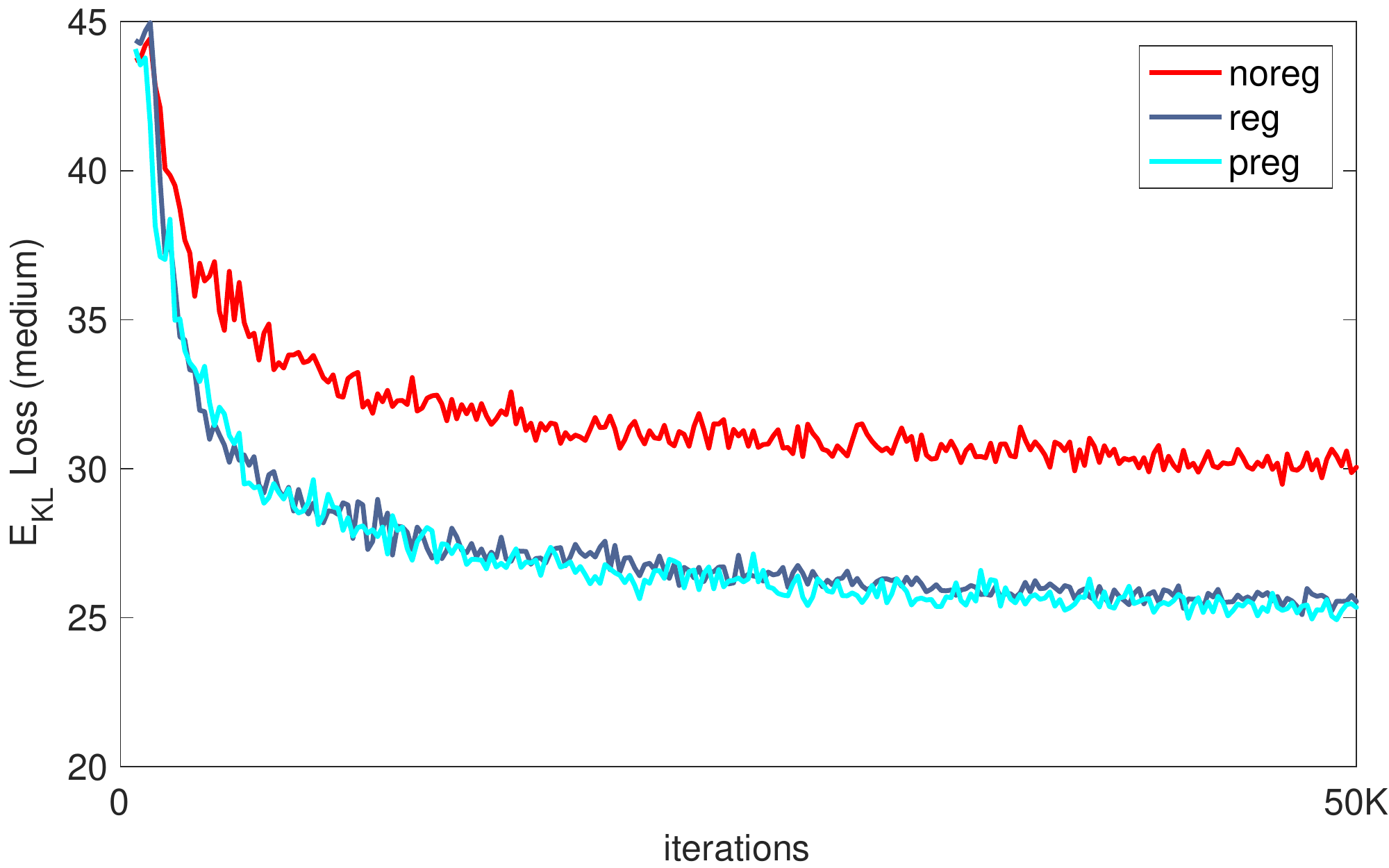}%
  }\\
\subfloat[triplets of (input, "noreg" palette, "patchreg" palette)]{%
  \includegraphics[width=0.99\columnwidth]{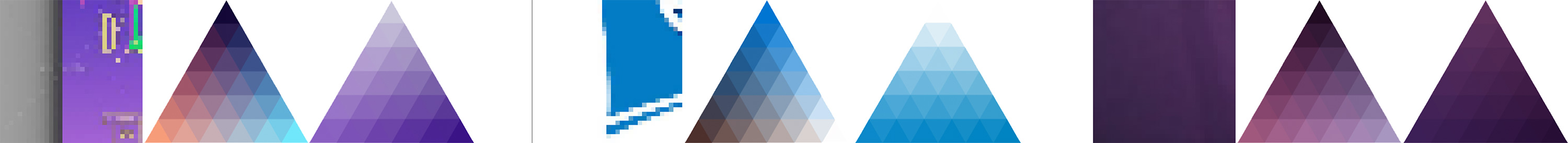}%
 }
\caption{\textbf{Palette Learning regularization:} "noreg" - unregularized learning, "reg" - regularized learning with histogram in $E_{KL}$ computed over the entire patch, "patchreg" - regularization with histogram in $E_{KL}$ computed over even smaller subpatches.  }
\label{fig:patchreg}
\end{figure}
We train and test our palette network (\S\ref{sec:pg}) on randomly selected image patches, with datasets carefully selected (\S\ref{ssec:datasets}) to match our goal of accurately approximating color distributions in regions of various categories of art and design. We experimented with different palette regularization schemes, and ultimately settled on a sum of $E_{L2}$ reconstruction loss and $E_{KL}$ divergence regularization with a weight of $0.0001$. We empirically found that patch-based histogram computation within a patch image for $E_{KL}$ computation (See discussion after Eq.~\ref{eq:e_kl}) produced slightly better results visually, but this was not reflected in our convergence graphs, as the metric itself is affected by this ("patchreg" in Fig.~\ref{fig:patchreg} for patch-based histogram, vs. regular histogram computation "reg"). This $E_{KL}$ regularization improves the $E_{KL}$ metric (Fig.~\ref{fig:patchreg}b) and encourages the encoded palettes to contain fewer unrelated colors (Fig.~\ref{fig:patchreg}c), while having no adverse effect on the reconstruction loss convergence (Fig.~\ref{fig:patchreg}a) . This regularization is key to both mapping estimation for color sail editing (\S\ref{ssec:pg_interact}), and region learning (\S\ref{sec:alpha_net}), where the greedy reconstruction in Eq.~\ref{eq:recon} could make use of unrelated colors in un-regularized palettes, derailing region learning.

\begin{figure}[ht]
\centering
\subfloat[$R_{\%}$ results on art types]{%
  \includegraphics[width=0.47\columnwidth]{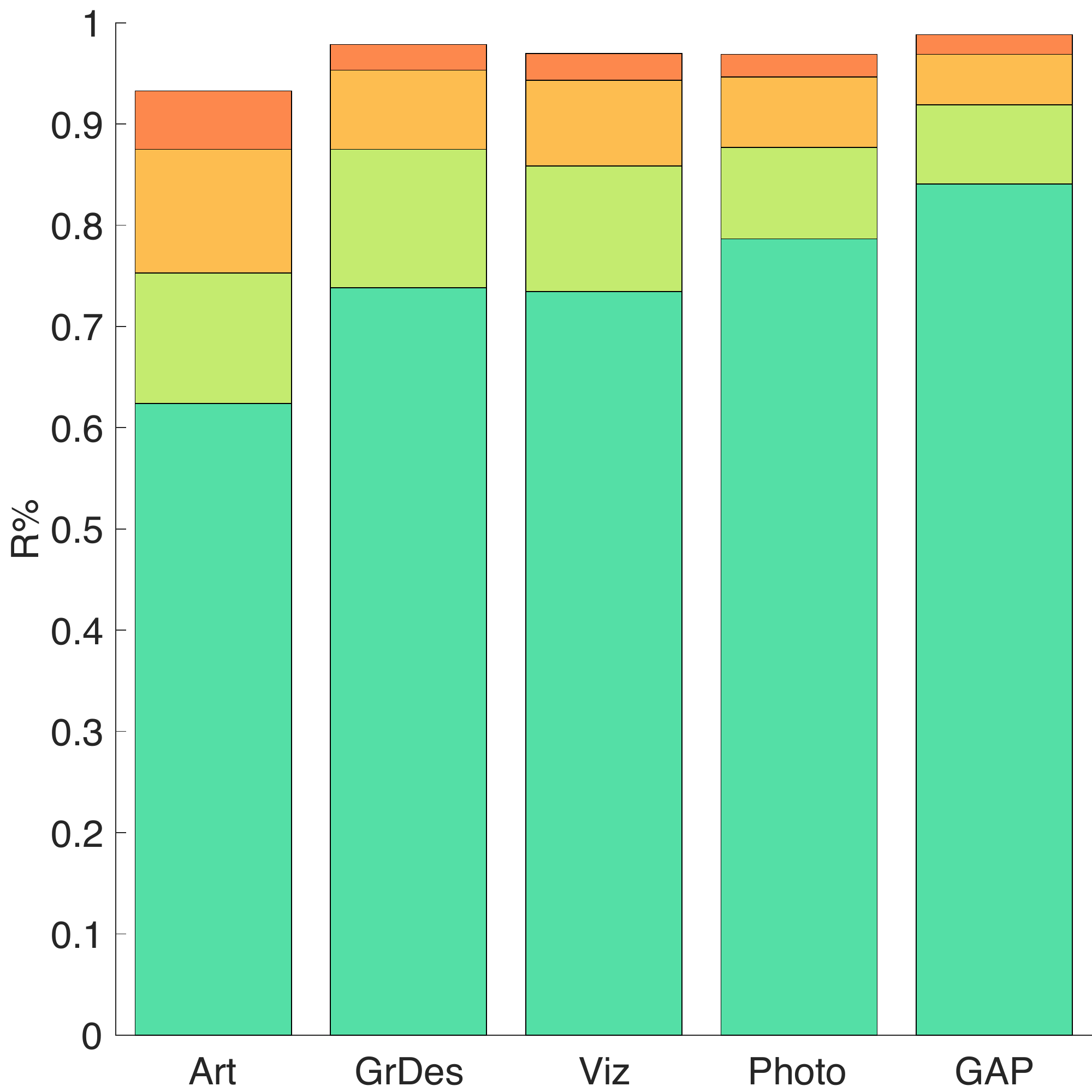}%
  }
\subfloat[$R_{\%}$ results on hardness]{%
  \includegraphics[width=0.47\columnwidth]{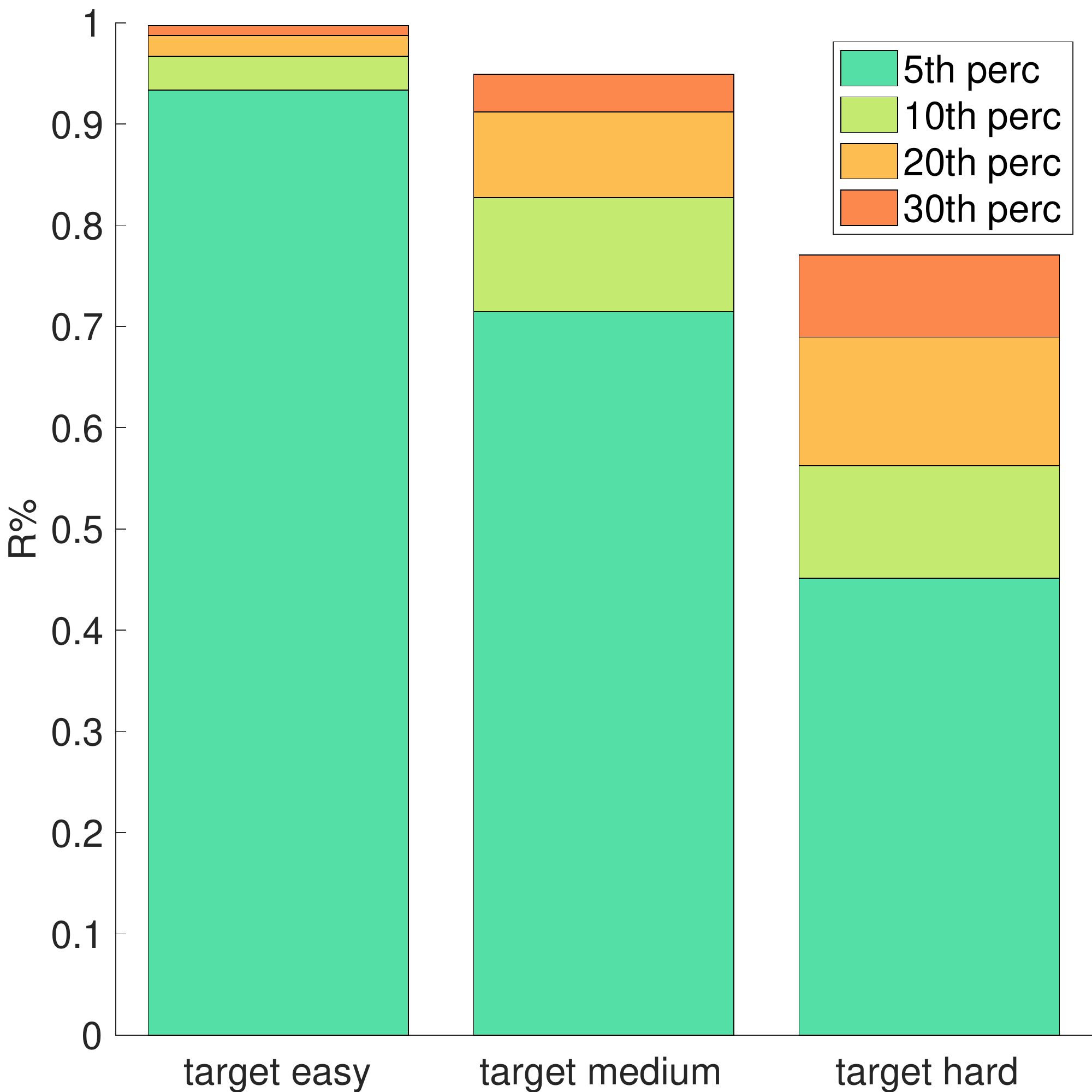}%
  }\\
  \subfloat[Maximal barely noticeable difference in $R_{\%}$ (we use $\delta=10$).]{%
  \includegraphics[width=0.99\columnwidth]{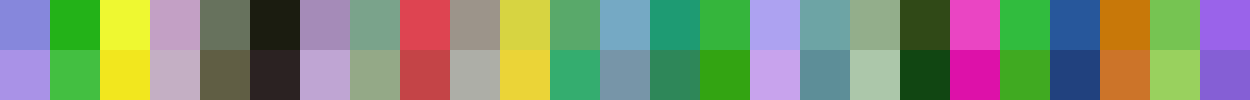}%
 }\\
  \subfloat[Results on challenging patches]{%
  \includegraphics[width=0.99\columnwidth]{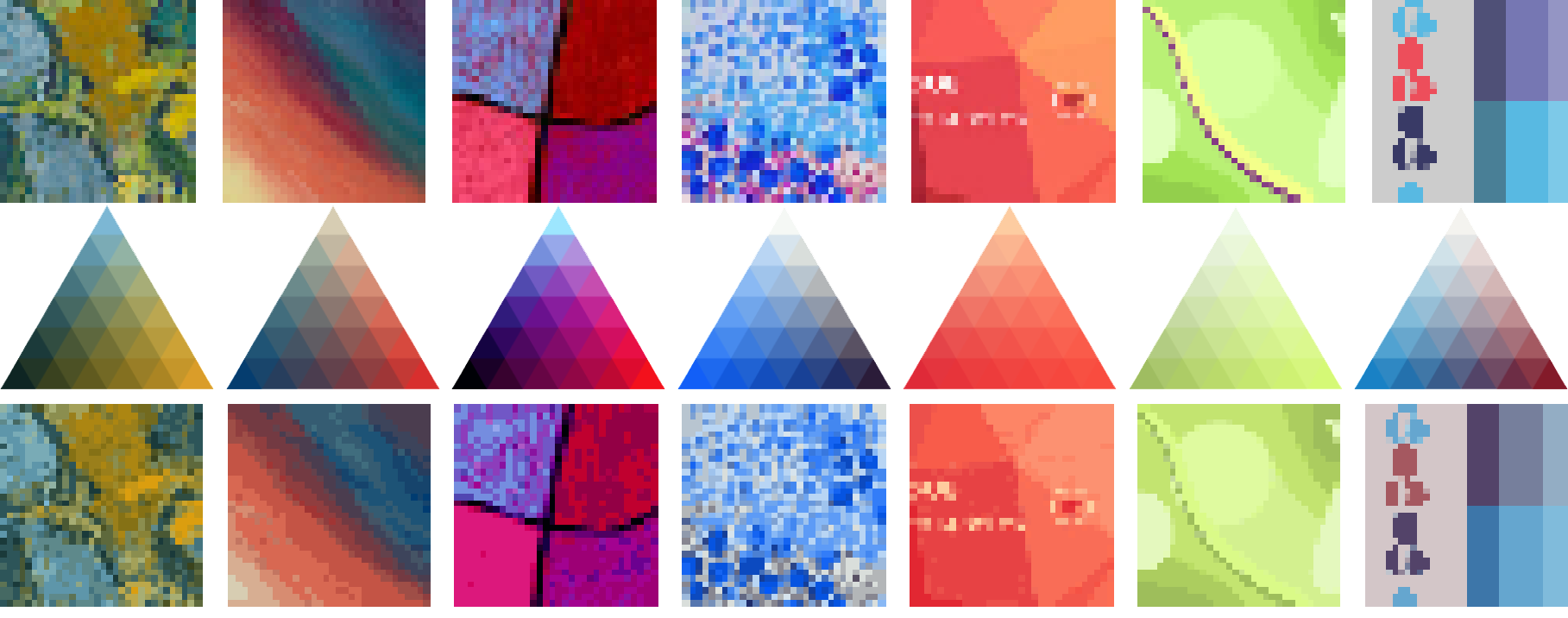}%
 }
\caption{\textbf{Palette Learning Performance:} reconstruction quality $R_{\%}$ (Eq.~\ref{eq:e_perc}) plotted
for art domains (a) and patch hardness levels based on entropy (b) (See \S\ref{ssec:datasets}.),
with maximal barely noticeable difference in $R_{\%}$ visualized in (c). Results on particularly difficult patches in (d), top to bottom: input patch, encoded palette, image reconstructed with one best matching palette color. }
\label{fig:patch_res}
\end{figure}

Our palette net achieves excellent performance on our target reconstruction quality $R_{\%}$ (Eq.~\ref{eq:e_perc}), with only 5\% of easy patches having $R_{\%}$ less than $0.93$, i.e. in 95\% of easy patches at least 93\% of pixels are reconstructed to be barely noticeable from the original. We find this percentile metric most informative, as we do not aim to perform well on all the patches -- some patches fall on the boundaries of regions we would like to model with color sails. The performance on medium patches is likewise good, with only 10\% of test patches having fewer than 83\% of pixels reconstructed sufficiently well. See Fig.~\ref{fig:patch_res} for percentile results on various art domains, hardness levels, as well as for a visualization of what we mean by "barely noticeable difference" and sample reconstructions.

\subsection{Alpha Mask Results}\label{ssec:alpha_eval}
\begin{figure*}[t!]
  \centering
  \includegraphics[width=7.2in]{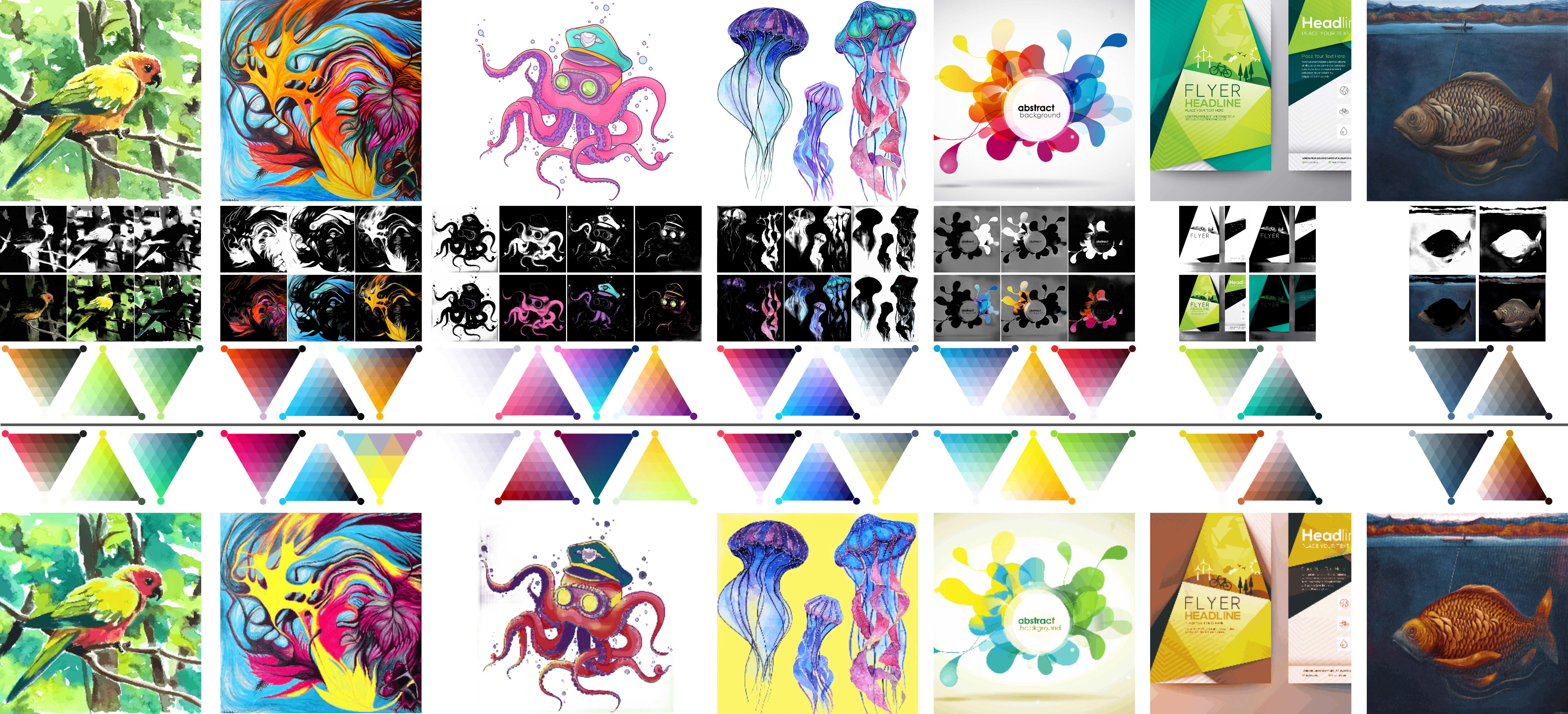}
  \vspace{-5mm}
  \caption{\textbf{Alpha results:} Top Half: Full color sail rig prediction results, Bottom Half: Recoloring results with color sail manipulation, with third and sixth results taken from our user study (\S\ref{sec:user_study}). Second and last column, artwork \copyright Maria Shugrina. }
  \label{fig:alpha_res}
\end{figure*}

\begin{figure}[ht]
\centering
\includegraphics[width=0.99\columnwidth]{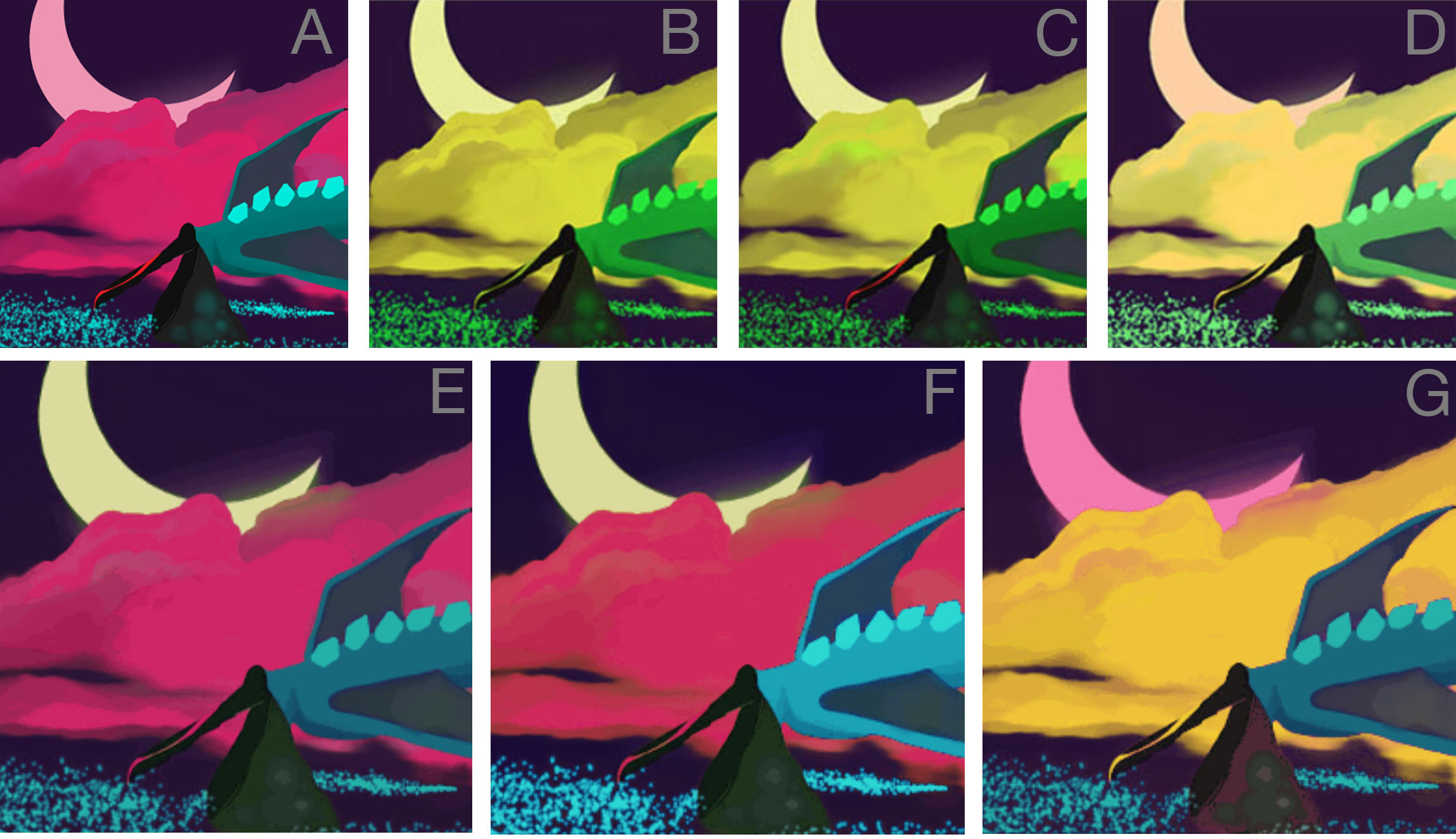}%

\caption{\textbf{Data-driven \st{semantic} vs. color based editing:}  Top Row: Recoloring using methods of ~\cite{aksoy2017unmixing}, \cite{tan2017decomposing} and \cite{chang2015palette}, as reported in Aksoy et al.'s supplemental material. Bottom Row: Recoloring using our model. Note how these methods relying solely on constant color cues recolor the moon together with the foreground region, but our alpha masks are able to isolate the moon. \edits{This shows promise of content-aware data-driven methods for image segmentation. Artwork \copyright Karl Northfell.}}
\label{fig:vs_segment}
\end{figure}

We show results from our joint alpha and color sail prediction in Fig.~\ref{fig:alpha_res}. In the top half, the rows shows the original image, the predicted alphas as binary masks and as mattes applied to the image, and the predicted color sail rig in order. In the bottom half, we demonstrate image recoloring after user interaction with the color sails, where the mapping is determined by our predicted alphas. The results demonstrate various aspects of versatility for recoloring using our system. The first image shows a simple recoloring with changes to the color of the sail vertices. The second column and third column demonstrate the effect of changing the patchwork level along with the colors to get more discrete or continuous color distributions respectively. The fourth column demonstrates the value of soft alphas for recoloring a non-trivial background seamlessly. \st{precision in our alphas that make it easy to recolor the non-trivial background area.} The other images show the versatility of our model, able to handle diverse domains such as poster designs and paintings. 

\edits{\subsubsection{Limitations and Future Directions}
While we observe that our alpha network can produce usable color rigs, we also see that it is not perfect. For example, the mouth of the fish is erroneously grouped with the surrounding sea (Fig.~\ref{fig:alpha_res}, last column), blended alphas can make editing difficult (Fig.~\ref{fig:alpha_res}, 5th column), and noise in the alphas can introduce artifacts (Fig.~\ref{fig:alpha_res}, 3rd column). It is worthwhile to note that our model theoretically supports user alteration of the alphas and could re-predict color sail mapping given user-provided or corrected alphas. This step is very fast, and could be used in an interactive loop (See \S\ref{sec:ux}).}

\edits{More importantly, we believe that our Deep Learning approach is only the first step toward learning user controls specifically optimized for color editing, given \emph{the context of a particular image}. By learning from color usage and color relationships in existing designs, our work is taking a new direction, when compared to prior work on color segmentation. }

\subsubsection{Comparison to Color Segmentation}
Unlike a large body of color segmentation work, we model each region not as a solid color, but as a wide distribution of colors. Representing each color alpha as a color sail allows additional creative possibilities of not only editing the global color characteristics within the alpha region, but the blending behavior as well. Moreover, because of our data driven approach, our alphas are more semantically related, as the model uses regularities in the data it has assimilated while learning along with the characteristics of the image at hand to produce a color sail rig. Fig.\ref{fig:vs_segment} shows recoloring results from~\cite{aksoy2017unmixing} in the top row vs. our editing in the bottom row \edits{(We cannot directly compare our results to the concurrent work of \cite{aksoy2018semantic}, as their segmentation model is trained to work on photographs, which is not our target domain of interest)}. Note how even though the moon and the clouds have the same color in the original image (A), our alphas allow separate editing (bottom row), where recent color based approaches do not (top row). \edits{While we cannot quantitatively conclude that our model selects more semantically meaningful regions than these approaches, we believe that it shows a lot of promise for segmentation that is aware of both the content and color relationships in an artistic image.}

\section{User Study}\label{sec:user_study}

\begin{figure}[h!]
\centering
  \includegraphics[width=0.95\columnwidth]{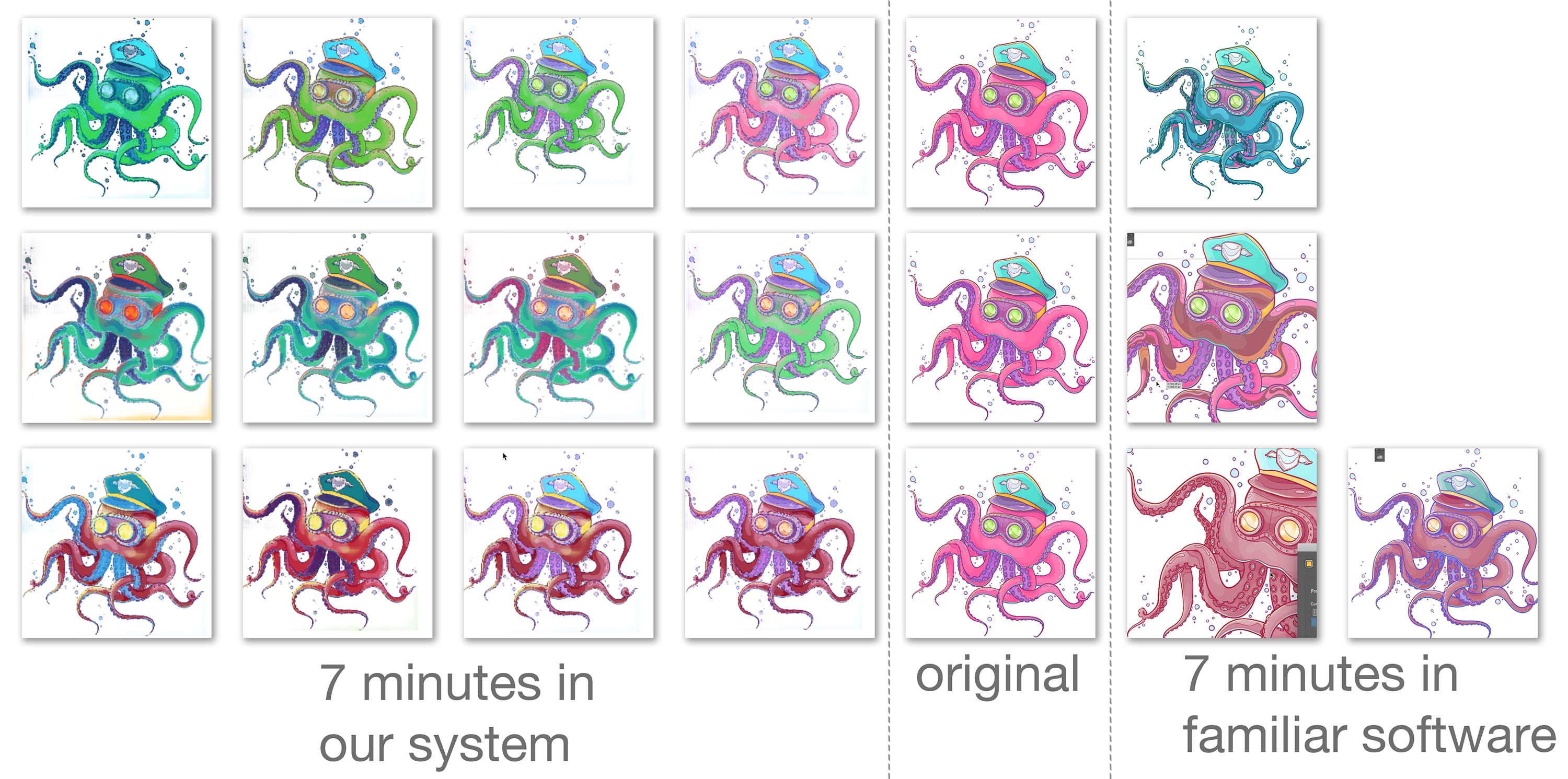}%
\caption{\textbf{User study:} Most participants were able to create many more quick variations using a color sail rig they were not familiar with than desktop applications they were intimately familiar with in the allotted 7 minutes. \edits{This figure shows color explorations of 3 users, using Adobe Illustrator and a layered vector octopus artwork, compared to using our system with automatically computed color sail rigs.}}
\label{fig:ustudy}
\end{figure}

The ultimate testament to our system is how useful it is to designers and artists in their real world workflow. We recruited 3 professional graphic designers, 2 professional product/ux designers, 2 professional or advanced digital illustrators, 2 traditional artists, and 2 people regularly using graphic design software for their work. We interviewed every participant about their use of color, and asked them to use our color sail interface, as well as a desktop application they were most familiar with for a color exploration task.

\textbf{Interview:} To understand if color exploration is a real need, we asked each subject whether  they create multiple versions of their designs specifically to visualize different color choices, not other design elements. 6 out of 11 emphasized that they do create different versions specifically for color exploration, while 3 reported that they create versions, but usually do not focus solely on color (versions do not really apply to 2 traditional artists). Several users stressed that they can only create multiple versions, \emph{if there is time}, implying the time commitment necessary to explore different color variations in existing software. When asked if they typically settle on a color scheme early on or continue to refine throughout the process, all but one said that color choice is an iterative process and changes throughout the design process (the one subject who replied otherwise works with company-defined branding color schemes).  

\textbf{Recoloring Task:} We asked each participant to adjust color choices in one of 3 pre-selected designs according to a loose inspiration (e.g. adjust this flyer for the fall season) using desktop software they were most comfortable with \footnote{\edits{For users electing to work with Adobe Illustartor, Illustrator layers were provided, simulating a realistic use case of adjusting a vector design.}} and using our web-based user interface with automatically computed color sail rigs. Both tasks were allocated 7 minutes, and the ordering was randomized. The intention of this task was not to produce a final result, but to visualize an alternative color choice to gain design intuition. In Fig.~\ref{fig:ustudy} we show the number of different explorations three Adobe Illustrator users were able to create for the same design using our unfamiliar software, and the tool they know very well. We believe that color sail based color exploration can augment rather than replace existing workflows designed for producing polished final results, but not for design exploration. In the questionnaire at the end of the study 8 out of 11 participants responded that this tool would complement the tools they currently use in their work (5 strongly agree, 3 agree on a 5-scale Likert scale), and 8 out of 10 felt they could be more efficient in their work if they had access to this system (4 strongly agree, 4 agree). Users also highlighted limitations of our work, such as \edits{difficulty in predicting the behavior of} the wind parameter, but it did not prevent them from experimenting with it. Most users made heavy use of the focus point parameter, because it has a very natural interpretation. Finally, a serious limitation of our current interface is inability to adjust the automatic alpha masks, or to visualize which areas of the region are affected by which color sail vertex. Despite these limitations, we believe that the number of quick color explorations our users were able to create using our system and their responses show that this is a promising research direction that can be useful to a wide variety of visual domains.


\section{Discussion}\label{sec:discuss}
\edits{Interviews with our user study participants confirm that exploring color variations is an important aspect of the artistic design process, regardless of the medium: illustration, graphic design or photography. While our alpha masks are not perfect for final hi-resolution image editing, they already allow rapid ideation through color sail interaction. Artists can use our method to explore quick variations before committing to a particular direction in their software of choice. On the interactive interface, there are two important avenues for future work: one, the coupling of color gamuts between color-sails in a rig, for example multiple color sails could share a vertex color or wind value; and two, sail tearing or the separation of a sail and its corresponding regions based on user feedback.}

\edits{While our alpha masks are not perfect, data-driven image segmentation is an important step toward truly semantics-aware image segmentation for editing. Using color cues alone to segment the image is inherently limiting (Fig.~\ref{fig:vs_segment}), and our approach shows promise for incorporating context and learning from data to produce masks specific to the context of a given artwork, as well as a specific editing method. We believe that we have only scratched the surface of this line of work, and hope that color sail representation can prove useful for continued innovation in this area. }

 
\edits{\emph{Color sails} are a novel artist-centric representation of color gamuts that combines the sparse simplicity of a discrete palette with the parametric power and color range of a continuous  representation. In addition to proposing a new representation this paper makes several contributions: we show that a color-sail interface is both usable and of utility to artists, i.e. is easy to control and understand, and enables common color-editing tasks that can be cumbersome using existing color tools, and show how color-sails can be used as a building block in conjunction with deep learning to infer and represent color for a wide range of input imagery. }


\edits{We believe that the color gamuts used in design today are strongly biased by current color representations and tools and that research into new representations and interaction with color, like color sails, will result in new forms of creative imagery.}

\section*{Acknowledgements}
We would like to thank our user study participants for their honest feedback and insights. We are also grateful to Nicole Sultanum for help with the user study design, and to Chris De Paoli for discussion.


\bibliographystyle{ACM-Reference-Format}

\bibliography{paper}

\appendix

\section{Technical Details}

\subsection{Interpolation}\label{app:interp}
There are many ways to implement Eq.~\ref{eq:pts} in practice, and our choices are designed to avoid problems\footnote{Certain computations cause numerical issues during gradient descent. For instance, normalizing the normal $\vec{n}$ would introduce a square root, which can destabilize derivatives near $0$ in some machine learning packages.} during  gradient-based machine learning (\S\ref{sec:learning}, \S\ref{sec:pg}). First, we do not normalize the normal $\vec{n}$. Because triangular color sails will have very different scales, depending on the distance between the color vertices, unscaled normal $\vec{n}$ provides a natural way of scaling the maximal "bending" by the triangle area. Second, rather than computing $d^2_{ijk}$ in absolute RGB and then trying to adjust the parameters of the falloff function based on triangle scale, we instead express $d^2_{ijk}$ as the distance between barycentric coordinates $\bm{u}_{ijk}$ and $\bm{u}_{111}$. The falloff function $f$ is then defined as a Gaussian with two parameters:
\begin{equation}
f(d^2) = \beta e^{-d^2/\alpha}	
\end{equation}
where $\beta$ simply scales the maximal wind displacement in RGB,
and $\alpha$ specifies the falloff variance (we use $\alpha=0.8$, $\beta=0.25$). The wind $w$ is in the range $[-1,1]$.

For completeness, the barycentric coordinates in Eq.~\ref{eq:pts} are defined as: 
	$\bm{u}_{300} = (1,0,0)$,
	$\bm{u}_{030} = (0,1,1)$,
	$\bm{u}_{003} = (0,0,1)$,
	$\bm{u}_{120} = (p_u, 1-p_u, 0)$,
	$\bm{u}_{210} = (1-p_v, v, 0)$,
	$\bm{u}_{102} = (p_u, 0, 1-u)$,
	$\bm{u}_{201} = (p_u + p_v, 0, 1-p_u-p_v)$,
	$\bm{u}_{012} = (0, p_v, 1 - p_v)$,
	$\bm{u}_{021} = (0, p_u+p_v, 1 - p_u - p_v)$. This also allows expressing $d^2_{ijk}$ cleanly in terms of squares of $p_u$ and $p_v$, which works well during gradient descent.
	
\subsection{Palette Network Structure}
For the palette network, we use input patches of size 32x32. A 10x10x10 normalized histogram of RGB colors is created and flattened before passing it through 4 fully connected layers. The fully connected layers have sizes of (700, 200, 50, $N_p$), where $N_p$ represents the number of parameters of our color sail encoding (\S\ref{sec:rep}). For the smaller patches used in Eq.~\ref{eq:e_kl}, we use patches of size 8x8. 

\subsection{Alpha Network Structure}
We use input images of size 256x256. Since the model is fully convolutional, inference is  possible on any input size. Our alpha network consists of a U-Net, where the encoder has 7 convolutional layers. Each convolutional layer employs 4x4 filters with stride 2 and uses the LeakyReLU non-linearity. The number of filters per layer are (64, 128, 256, 256, 256, 256, 256). The decoder uses the same hyperparameters for transpose convolutions with U-Net like skip connections. The activations are batch normalized after every convolution.

\section{Datasets}\label{sec:data}
Our goal is to support creative color work for a wide range of 2D graphics, including painting, illustration, graphic design and visualization, calling for a representative dataset. To our knowledge, no existing dataset covers such a broad set of categories, and therefore we collect custom datasets. 

\subsection{Data Collection}\label{ssec:dcollect}
 We collected images in the following broad set of categories: 1) art, which includes painting and illustration, 2) graphic design, which includes more targeted use cases, such as poster, logo and user interface design, 3) visualization, where primary goal is to graphically show information. We include photography for reference, but are not targeting this visual style in our work, as editing natural images calls for a different set of tools. For physical painting, we used the Google Search API~\cite{googlesearch} to collect 100 images by each of 30 well-known artists. For digital painting, we queried for "digital painting" with a category "Painting" on \url{behance.net} \cite{behance}. For "illustration", we collected images from the curated "Illustration" gallery on behance.net, and also collected images of New Yorker cover art~\cite{newyorker}, which spans a wide range of illustration styles. For poster graphic design, we searched pinterest.com~\cite{pinterest} for "poster design". For logo design, we searched designinspiration.com \cite{designinsp} for "logo design", and also collected logos of the Fortune 500 companies~\cite{fortune500}. For user interface design, we focused on mobile apps and searched pinterest.com for "app ux design" and collected screenshots from mobile-patterns.com \cite{mobilepatterns}.  For visualization, we used an existing MassViz dataset of hand-labeled images~\cite{borkin2013makes}. Finally, for high-quality professional photography we used images from the curated "Photography" gallery on behance.net. These datasets are summarized in Table~\ref{tb:data}.
 
 \begin{table}[ht]
\begin{center}
\small{
\begin{tabular}{@{}c|l|l|c|c@{}}
\bf{Category} & \bf{Subcat.} & \bf{Src} & \bf{\#Total.}  & \bf{\#Test}  \\
\Xhline{1.3pt}
\multirow{3}{*}{\bf{Art}}&(A1) Phys.\ paint.&GS&2808&\multirow{3}{*}{600}\\
\cline{2-4}
&(A2) Digital paint.&BH+DI&1075+1011&\\
\cline{2-4}
&(A3) Illustration&BH+NY&1046+2497&\\
\Xhline{0.8pt}
\multirow{3}{*}{\shortstack{\bf{Graphic} \\ \bf{Design}}}&(G1) Posters&PN&988&\multirow{3}{*}{600}\\
\cline{2-4}
&(G2) Logos&DI+GS&1048+500&\\
\cline{2-4}
&(G3) UX design&PN+MP&933+1394&\\
\Xhline{0.8pt}
\bf{Viz.}&(V1) Misc.~viz.&MassViz&2000&300\\
\Xhline{0.8pt}
\bf{Photo}&(P1) Misc.~photo&BH&1078&300\\[3mm]
\Xhline{2.0pt}
\bf{Uncategorized}&(M1) Misc.&GAP&165.5K&3000\\
\Xhline{1.3pt}
\end{tabular}
}
\end{center}
\caption{\textbf{Image Datasets}: Source legend: BH - behance.net, DI - designspiration.net, GS - Google Search API, MP - mobile-patterns.com, MV - MassViz dataset~\cite{borkin2013makes}, NY - New Yorker covers from newyorker.com via Google Search API, PN - pinterest.com, GAP - Google Art Project. }
\label{tb:data}
\end{table} 

 While these hand-picked datasets cover a wide range of color usage for palette learning, they are too small for alpha mask learning. To augment our dataset, we additionally collect a larger uncategorized dataset by crawling Google Art Project(GAP)~\cite{googleart} for artists, movements and a subset of the mediums. All the datasets collected are summarized in Tb.~\ref{tb:data}. Because the GAP data has a very different distribution from our target data, care must be taken to form representative test and train datasets.

%


\subsection{Datasets} \label{ssec:datasets}

\begin{figure}[ht]
\centering
\subfloat[example patch histogram entropy and image colorfulness values]{%
  \includegraphics[width=0.99\columnwidth]{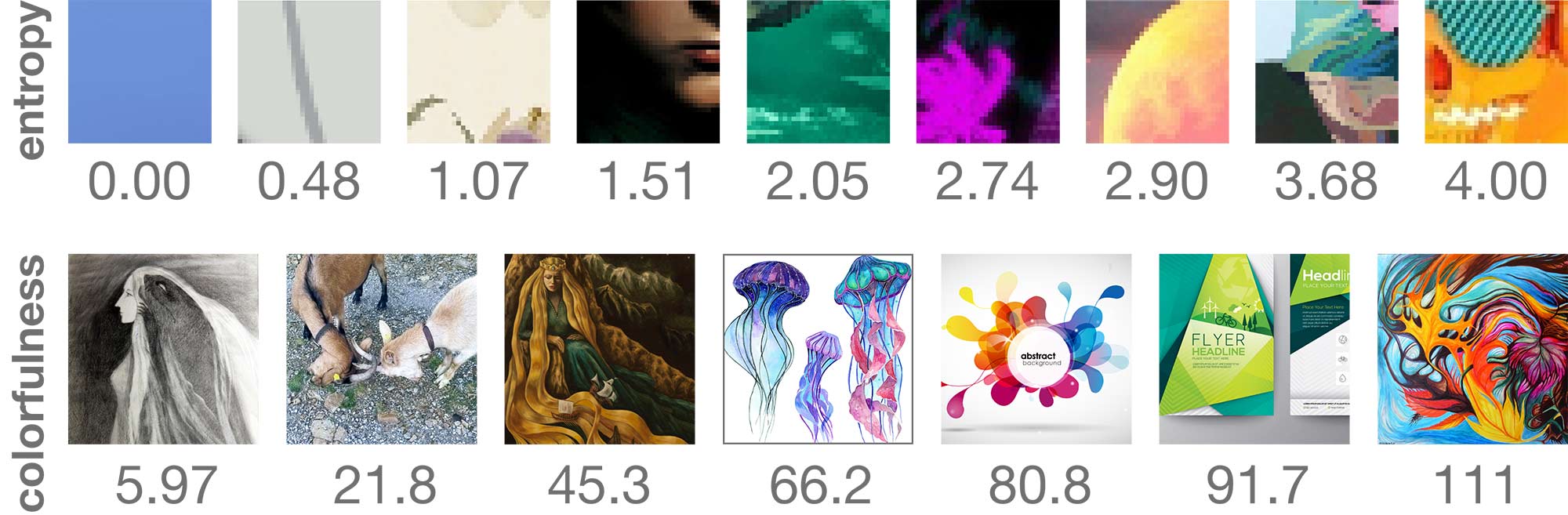}%
  }\\
\subfloat[colorfulness]{%
  \includegraphics[width=0.45\columnwidth]{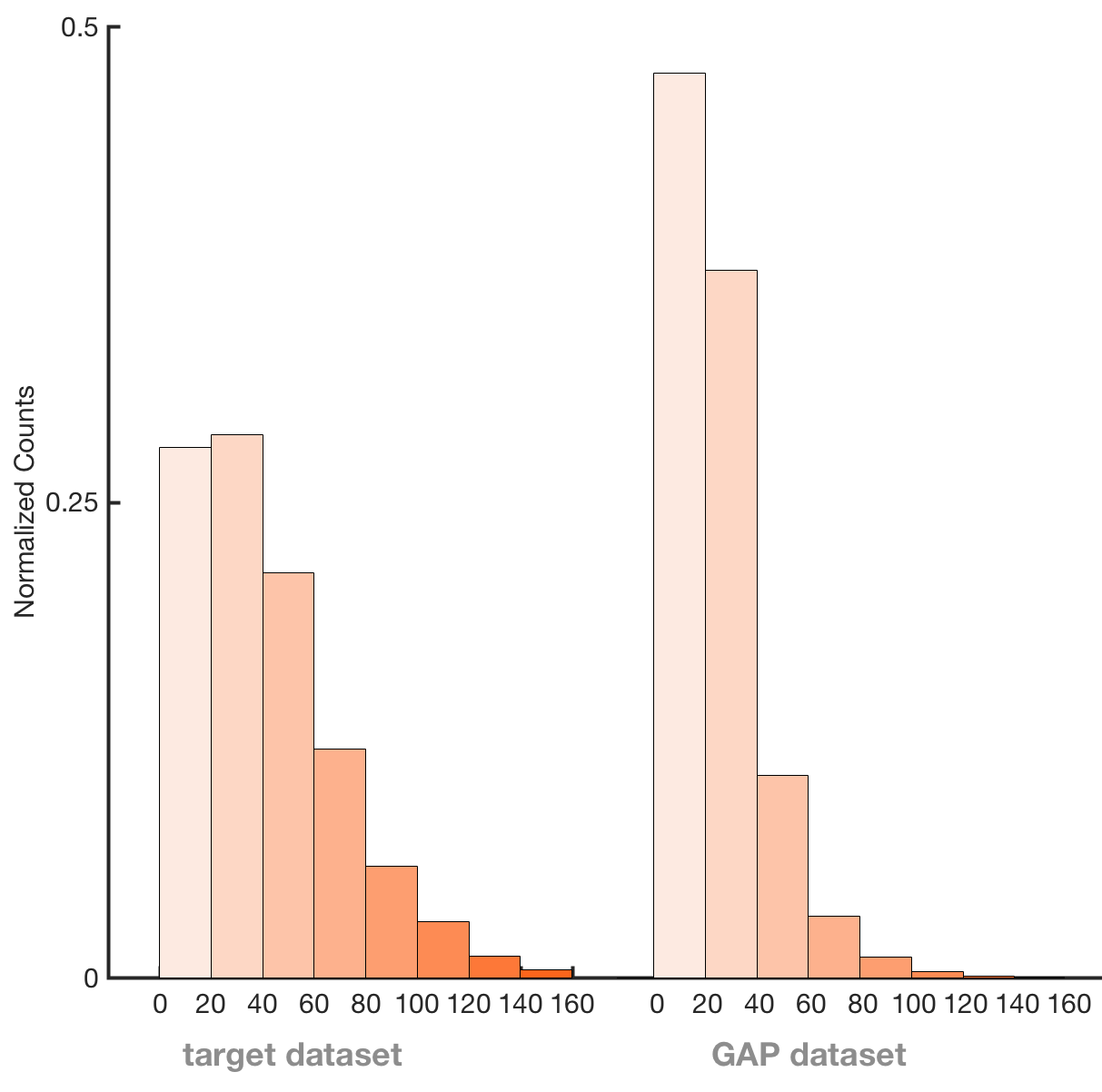}%
  }
\subfloat[patch entropy]{%
  \includegraphics[width=0.54\columnwidth]{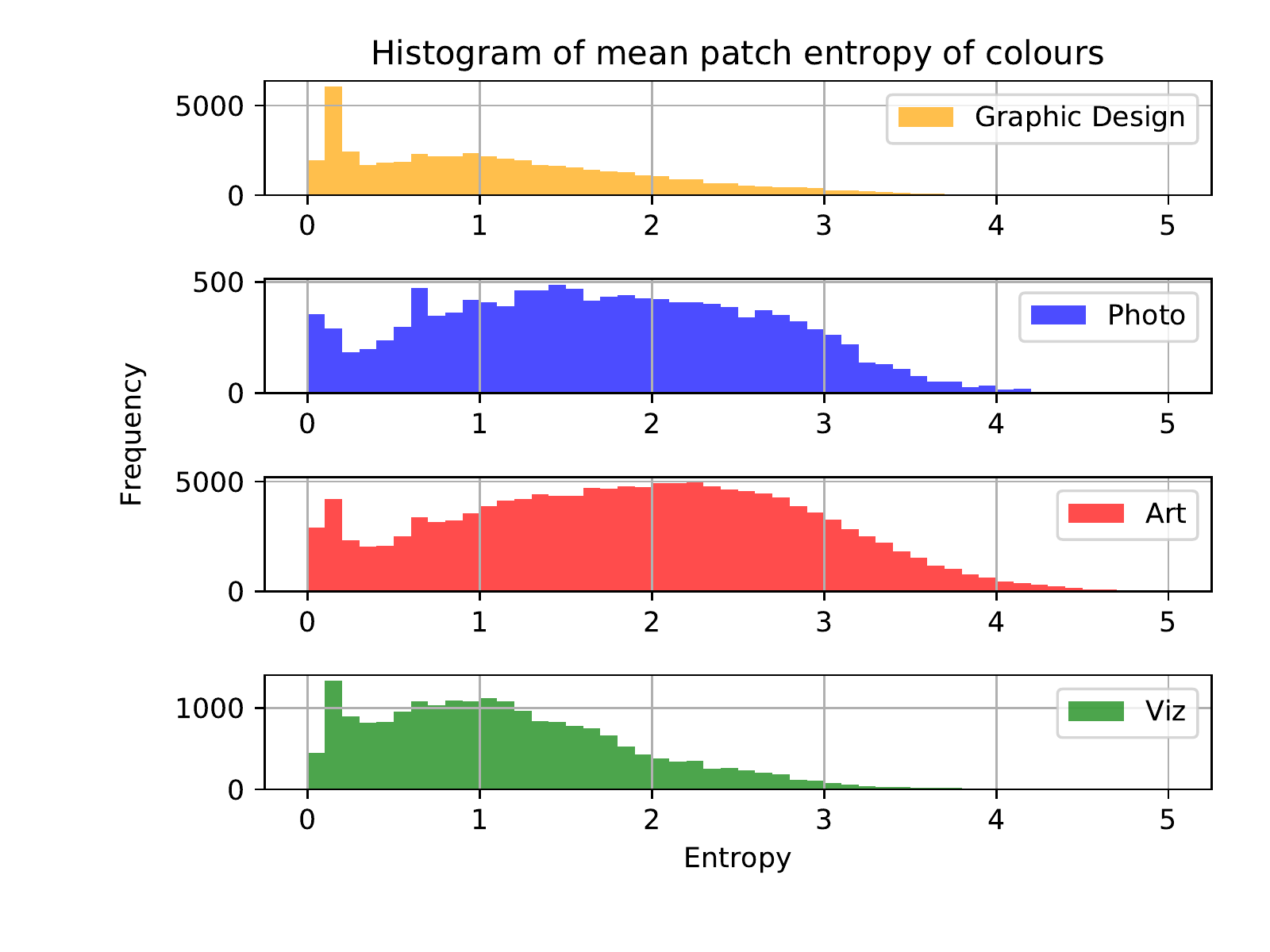}%
  }
\caption{\textbf{Data Analytics:} Sample entropy (a, first row), and colorfulness (a, second row)
of image patches and images, respectively. We use colorfulness to weigh our unlabeled dataset
(b) to better approximate target data distribution, and entropy to designate various levels of difficulty for palette learning test sets.}
\label{fig:datastats}
\end{figure}

We withhold randomly selected test data from different categories to create a test set that reflects the design disciplines our work is targeting (See Tb.~\ref{tb:data}, last column). We refer to this test set of 1800 images as the \textbf{target test set}. In addition, we withhold a larger \textbf{dirty test set} from uncategorized GAP data, which does not necessarily represent our target media (for example, ancient manuscripts, rough sketches, architecture). We also withhold small validation sets for all categories to monitor and tune training. We refer to the remaining categorized training data as the \textbf{target training set}, and remaining GAP data as the \textbf{dirty training set}.

\subsubsection{Image Train Sets} By browsing random images from our GAP test set, we discovered them to be on average much less colorful than the images in our target set, which could have a serious effect due to the central role of color in our networks. Using an approximation of the perceptual colorfulness metric~\cite{hasler2003measuring}, we confirmed that GAP images are less colorful than those in our target dataset (Fig.~\ref{fig:datastats}a, b). To correct for the discrepancy, we weigh GAP data in a way to make its colorfulness distribution equal to that of our target training set when selecting batches during stochastic gradient descent. Each batch for both palette and alpha learning consists of equal number of target and GAP samples.

\subsubsection{Patch Test Sets} We initially tried palette learning on random 32x32 patches extracted from images rescaled to 512x512 in the combined target and dirty training sets, but quickly learned that random patches mostly contain muted solid colors from background and low-frequency areas of the image, making it not only hard to train, but to evaluate performance on such data. To overcome this, we observe that histogram entropy computed over the patch colors is a rough indicator of color complexity present in the patch (Fig.~\ref{fig:datastats}a). By examining patches of various entropy levels, we designated "easy" patches with entropy $<1.5$ as easy, patches with entropy over $3$ as "hard", and the rest as "medium", and split our test set accordingly to help us better evaluate performance. See Fig.~\ref{fig:datastats} for patch entropy distributions in our datasets. 

\subsubsection{Patch Train Sets} In order to encourage more difficult patches to occur in the training set during target patch selection, we center bias our random patch selector, as the central area of the artwork is more likely to contain the more detailed main subject. In addition, we found it better to generate patches of random size and then rescaling the distribution. This makes the training set of the palette network more analogous to the region-based input it sees during alpha network training (we plan to address further palette network tuning and exploration in future work).

\end{document}